\documentclass{article}
\usepackage[english]{babel}
\usepackage{amsmath}
\usepackage{graphicx}
\usepackage{hyperref}
\usepackage{caption}
\usepackage{subcaption}
\usepackage{authblk}
\usepackage[utf8]{inputenc} 
\usepackage[dvips]{geometry}
\usepackage{indentfirst}
\usepackage{relsize,exscale}
\usepackage{amsfonts}
\usepackage{amsmath}
\usepackage{epstopdf} 
\usepackage{amssymb}
\usepackage{commath} 
\usepackage[dvips]{epsfig} 
\usepackage{float} 
\usepackage{setspace}
\usepackage{titlesec, blindtext, color}
\usepackage{makeidx}
\usepackage{ulem} 
\usepackage{cancel}
\usepackage{appendix}
\usepackage{xcolor} 
\usepackage{braket}
\usepackage{cancel}
\usepackage{relsize}

\DeclareMathOperator{\csch}{csch}
\DeclareMathOperator{\sign}{sign}

\title{The Unruh effect under the de Broglie-Bohm perspective}
\author[1]{Matheus M. A. Paixão \footnote{Email: \href{mailto:matheuspaixao@cbpf.br}{matheuspaixao@cbpf.br}}}
\author[1]{Olesya Galkina \footnote{Email: \href{mailto:galkina@cbpf.br}{galkina@cbpf.br}}}
\author[1]{Nelson Pinto-Neto \footnote{Email: \href{mailto:nelsonpn@cbpf.br}{nelsonpn@cbpf.br}}}
\affil[1]{CBPF - Centro Brasileiro de Pesquisas F\'{\i}sicas, Rua Doutor Xavier Sigaud 150, 22290-180, Rio de Janeiro, Brazil}
\date{}
\begin{document}
\maketitle

\begin{abstract}
We investigate the Minkowski ground state associated with a real massless scalar field as seen by an accelerated observer under the perspective of the de Broglie-Bohm quantum theory. We use the Schrödinger picture to obtain the wave functional associated with the Minkowski vacuum in Rindler coordinates, and we calculate the field trajectories through the Bohmian guidance equations. The Unruh temperature naturally emerges from the calculus of the average energy, but the Bohmian approach precisely distinguishes between its quantum and classical components, showing that they periodically interchange their roles as the dominant cause for the temperature effects, with abrupt jumps in the infrared regime. We also compute the power spectra, and we exhibit a very special Bohmian field configuration with remarkable physical properties. 
\end{abstract}

\section{Introduction}
The construction of a consensual quantum theory of gravity is one of the toughest and most intriguing challenges of theoretical physics. One of the attempts to understand how quantum effects appear in gravity theories is to consider Quantum Field Theory (QFT) in curved spacetimes, where important phenomena such as Hawking radiation and cosmological particle production appear. It should be noted that even in flat spacetime the particle concept is, in general, observer-dependent, as it is well-known from the Unruh effect\cite{Fulling, Davies, Unruh}. This is particularly important for understanding particle emissions from black holes\cite{Hawking1, Hawking2}, once we achieve similar results with much simpler calculations.

A simple and didactic way to address the Unruh effect is to consider a free scalar field in a flat $2$-dimensional space from two perspectives: for an inertial observer in Minkowski space, and a uniformly accelerated observer with respect to the first one. According to QFT, both disagree on the number of particles. In the Minkowski ground state, the number of particles in the inertial frame is zero, while for the accelerated (Rindler) observer there are particles in a Bose-Einstein distribution with temperature proportional to the acceleration,

\begin{equation}
    T=\frac{\hslash a}{2\pi \kappa_b c}.\label{UnruhTemp}
\end{equation}
The key point here is that the inertial observer's vacuum state differs from the vacuum of a non-inertial one. Therefore, the number of particles defined in Rindler space concerning the inertial vacuum is different from zero\footnote{A similar result can be obtained if we consider fermions instead of a scalar field. In this case, we obtain a Fermi-Dirac distribution with the same temperature\cite{fermionic unruh effect, fermionic unruh effect2}.}. Despite being an interesting phenomenon, its observation is quite challenging. As a rough estimate, to reach a temperature of $1K$ an acceleration of $10^{19}m/s^2$ is necessary. Nevertheless, experimental observations of the Unruh effect are discussed in\cite{Rosu,Hu,Akhmedov,Martinez,Gooding-Unruh}. For Hawking radiation, see\cite{Kolobov, Nova, Steinhauer, Balbinot, Belgiorno}.

Usually, the Unruh temperature Eq.~\eqref{UnruhTemp} is obtained in the Heisenberg picture in the framework of the standard probabilistic view of quantum theory. However, this standard approach cannot be extended to a unified quantum picture of the Universe, as it assumes the necessity of a classical world outside the quantum system, where measurements are realized and definite outcomes are obtained~\cite{Omnes}. As in quantum cosmology the quantum system is the whole Universe, there is no place for this classical domain, and the standard approach cannot be applied.
In this sense, the de Broglie-Bohm approach to quantum theory (dBB)\cite{deBroglie, Bohm1, Bohm2} is very appropriate\footnote{There are other alternative possibilities, like the many world interpretation and collapse models, ~\cite{mwi,collapse}, but we will not use them in this paper.}. In this framework, point particle positions or field configurations are supposed to have objective reality, and their dynamics are dictated by the wave function through the so-called guidance equations. If the uncertainty of initial positions or field configurations is given by the Born rule at some initial time, then all probability predictions of quantum mechanics can be recovered. Hence, this formulation does not yield different experimental results as in usual quantum theory, but, besides opening the route to quantum cosmology, it gives new understandings of quantum phenomena which might be useful\footnote{If the uncertainty of initial conditions is not given by the Born rule in some initial time, then the theory can lead to different results from usual quantum mechanics up to the time when the Born rule is recovered, the so-called quantum equilibrium state. For details on this interesting possibility, see Ref.~\cite{Valentini1991I}.}. First of all, it solves the measurement problem \cite{nelson universe}, so there is no need for an external classical world for the description of quantum measurements. This allows, for example, the understanding of the quantum-to-classical transitions of cosmological perturbations \cite{Pinto-Neto:2013npa}. Secondly, the classical limit of quantum systems can be described in terms of the quantum potential $Q$, in such a way that when $Q$ approaches zero, the quantum trajectories approach the classical ones. Furthermore, in the dBB framework it is possible to study the quantum singularities of cosmological models in more detail and, consequently, to identify non-singular quantum models through the Bohmian solutions, such as bouncing models, where the Friedmann solution can be reproduced under certain limits \cite{bounce,bounce2,almeida}. Finally, the guidance equations for the universe wave function provide its time evolution, even though the quantum equations for $\Psi_{univ}$ do not admit a Schrödinger form. Schrödinger-like equations appear when we study subsystems, ensuring the usual time evolution and a probabilistic interpretation in terms of the Born rule~\cite{nelson universe}. 

In such an appealing scenario, it is valid to argue how the Unruh effect arising in Quantum Field Theory can be addressed under the de Broglie-Bohm's perspective. In order to obtain a guidance equation, we need to tackle this problem using Schrödinger's representation of the fields\cite{Hatfield}, in order to get the wave functional associated with the Minkowski vacuum in Rindler space. Using the results of Ref.~\cite{Freese}, we were able to obtain the complete wave functional solution, the guidance relations, and their integration, obtaining the Bohmian field trajectories. The Unruh temperature is obtained in this framework, and its origin as a quantum effect is discussed in detail. In particular, highly abrupt jumps between quantum and classical dominance, which can be discriminated only in the dBB quantum theory, happen periodically in the infrared limit, or in the high acceleration limit. We discuss whether this property can lead to experimental consequences. Also, as the Bohmian approach is manifestly non-local, even for relativistic quantum field theory \cite{BHK,BHK2,Kaloyerou,Kaloyerou2} (due to the appearance of the non-linear and non-local quantum potential), the entanglement between the left and right Rindler wedge fields \cite{Unruh-Wald} can be addressed within a different perspective.

This work is divided into a right-hand Rindler wedge analysis and a complete manifold analysis, which includes both right and left Rindler wedges. In section \ref{KG}, we summarize Klein-Gordon theory in the Rindler space, and we review some results regarding the wave functional of a massless scalar field in the right Rindler wedge. We then present the associated de Broglie-Bohm theory in subsection \ref{dbb}. In subsection \ref{Mean values}, we calculate the mean number of Rindler particles by computing the averages of the quantum and classical quantities. In the following subsections \ref{field trajectories rhw} and \ref{power spectrum rhw}, we obtain the Bohmian trajectories, with emphasis on a very peculiar particular one, we analyze the asymptotic expansions for the low and high acceleration regimes, and we calculate the power spectrum. In section \ref{complete analysis}, we extend our analysis to the complete spacetime, following the same order: we obtain mean values and Bohmian trajectories, analyze the asymptotic expansions and calculate the power spectrum. Finally, our conclusions are presented in the last section. Natural units are used throughout this work, with $c=\hbar=\kappa_b=1$.

\section{The wave functional approach and its Bohmian interpretation in the right-Rindler wedge}\label{KG}

In this section, we obtain the Minkowski wave functional in Rindler coordinates for the right wedge based on the work \cite{Freese}. The trajectory of an observer with constant acceleration $a$ and coordinates $(t,x)$ in some Minkowski frame is given in terms of the transformations
\begin{align}
\begin{tabular}{ l }
$x\left(\tau,\xi\right)=\dfrac{e^{a\xi}}{a}\cosh(a\tau)$\\
$\hspace{1.0mm}t\left(\tau,\xi\right)=\dfrac{e^{a\xi}}{a}\sinh(a\tau),$\\
\end{tabular}\label{xr}
\end{align}
where $-\infty < \tau < \infty$ and $-\infty < \xi < \infty$ are the Rindler coordinates\cite{Rindler1960, Rindler1966, Valdes, Matsas}. The horizons $t=\pm x$ are reached when $\tau \to \pm \infty$, while the origin is achieved when $\xi \to -\infty$. The line element $ds^2=-dt^2+dx^2$ can be rewritten as $ds^2=e^{2a\xi}(-d\tau^2+d\xi^2)$, implying a conformal invariance between the metrics, with a conformal factor $e^{a\xi}$.

Consider a real massless scalar field $\phi$ in Minkowski space described by the action
\begin{align}
\label{Maction}S=\frac{1}{2}\int dtdx \biggr\{\left(\frac{\partial \phi}{\partial t}\right)^2-\left(\frac{\partial \phi}{\partial x}\right)^2\biggl\}.
\end{align}
The field $\phi$ satisfies the Klein-Gordon equation $(-\partial_t^2+\partial_x^2) \phi = 0$. According to the transformations \eqref{xr}, the action \eqref{Maction} becomes 
\begin{align}
S=\frac{1}{2}\int d\tau d\xi \left(\left(\frac{\partial \phi}{\partial \tau}\right)^2-\left(\frac{\partial \phi}{\partial \xi}\right)^2\right).\label{Raction}
\end{align}
As a result, the associated equation of motion is $(-\partial_{\tau}^2+\partial_{\xi}^2) \phi = 0$, which is just the Klein-Gordon equation in Rindler coordinates.

However, there is a slight difference between the actions \eqref{Maction} and \eqref{Raction}. While $\phi$ can be defined all over Minkowski spacetime, the same is not valid in Rindler coordinates, where $\phi$ can be defined only in the right-hand wedge ($-x<t<x$ and $0<x<\infty$). The Fourier expansion in Minkowski modes in this region is given by
\begin{equation}\label{FourierM}
    \phi(t,x)=\sqrt{\frac{2}{\pi}}\int_{0}^{\infty} dk \sin(kx)\phi_k^M(t),
\end{equation}
where $(\phi_{k}^{M})^{*}=\phi_{k}^M$. On the other hand, the expansion in Rindler modes is
\begin{align}
\phi(\tau,\xi)=\int_{-\infty}^{\infty}\frac{dk'}{\sqrt{2\pi}}e^{ik'\xi}\phi_{k'}^R(\tau),\label{FourierR}
\end{align}
with $(\phi_{k'}^{R})^{*}=\phi_{-k'}^R$. 

From the action \eqref{Maction}, we obtain the Minkowski Hamiltonian 
\begin{align}\label{Hamiltonian M 2}
   H^M=\frac{1}{2}\int_{0}^{\infty} dk 
   \left(-\frac{\partial^2}{\partial^2\phi_{k}^{M}}+k^2\left( \phi^M_{k}\right)^2 \right).
\end{align}
Consider the following decomposition for the wave functional  
\begin{align}
\Psi[\phi,\eta]=\prod_{k>0} \Psi_{k}[\phi_k,\phi_k^{*},\eta], \label{psi decomposition}
\end{align}
where $\eta$ is a temporal variable. Each term $\Psi_k$ satisfies an independent Schrödinger equation 
\begin{equation}\label{Schrödinger eq M}
    i\frac{\partial\Psi^M_k[\phi_k^M,t]}{\partial t}=\frac{1}{2} \left(-\frac{\partial^2}{(\partial \phi^{M}_k)^2}+k^2 \left( \phi^M_{k}\right)^2 \right)\Psi^M_k[\phi_k^M,t],
\end{equation}
that admits as a ground state solution
\begin{align}
(\Psi_{k}^M)_0[\phi_k^M,t]=N_k\exp\left(-\frac{1}{2}k\left(\phi^{M}_k\right)^2-\frac{i}{2}kt \right).\label{ground state M}
\end{align}
Therefore, the wave functional \eqref{psi decomposition} becomes
\begin{align}
(\Psi^M)_0[\phi_k^M,t]=N\exp\left(-\frac{1}{2}\int_{-\infty}^{\infty}dk k\left(\phi^{M}_k\right)^2-i\Omega_0 t \right),\label{ground state M complete}
\end{align}
with $\Omega_0$ the zero-point energy and $N$ a normalization constant.

In order to obtain the Minkowski vacuum in Rindler coordinates, we need to write the $\phi_k^M$ modes in terms of $\phi_k^R$. By inverting the equation \eqref{FourierM} and using the expansion \eqref{FourierR}, we can write
\begin{align}
    \phi_k^M=\int_{-\infty}^{\infty}dk'A(k,k')\phi_{k'}^R,\quad k>0\label{phik expansion}
\end{align}
where the coefficient $A(k,k')$ can be calculated in their common spacelike hypersurface at $t=\tau=0$ (see Ref.~\cite{Freese} for details) and it is given by
\begin{align}
    A(k,k')=\frac{1}{a\pi}\Gamma\left(1+\frac{ik'}{a}\right)\cosh{\left(\frac{\pi k'}{2a}\right)}\left|\frac{k}{a}\right|^{-1-i\frac{k'}{a}}. \label{A(k,k')}
\end{align}
Therefore, substituting  equations \eqref{phik expansion} and \eqref{A(k,k')} into \eqref{ground state M complete}, we obtain the Minkowski ground state wave functional at $t=\tau=0$ in terms of the Rindler field configuration $\phi_{k}^R$
\begin{align}
    (\Psi^M)_0[\phi_{k}^R,\phi_{k}^{R*},0]=N\exp\left(-\int_{0}^{\infty} dk k \coth\left(\frac{\pi k}{2a}\right)\phi_{k}^R\phi_{k}^{R*} \right).
\end{align}
Due to the decomposition \eqref{psi decomposition} we have
\begin{align}
    (\Psi_{k}^M)_0[\phi_{k}^R,\phi_{k}^{R*},0]=N_{k}\exp\left(-k \coth\left(\frac{\pi k}{2a}\right)\phi_{k}^R\phi_{k}^{R*} \right).\label{psiM}
\end{align}

The Rindler ground state wave functional is obtained similarly. From the action \eqref{Raction}, we obtain the Hamiltonian
\begin{align}\label{Hamiltonian R 2}
   H^R=\int_{0}^{\infty} dk 
   \left(-\frac{\partial^2}{\partial\phi_{k}^{R*}\partial\phi_{k}^{R}}+k^2|\phi^R_{k}|^2 \right),
\end{align}
where we split the integral into two equal contributions for positive and negative values of $k$ and perform the change of variables $k\to-k$ in the negative part. As a consequence of this choice, the energy of each mode will be two times its original value. Therefore, adopting the decomposition \eqref{psi decomposition}, we have a Schrödinger equation for each $\Psi_{k}^R$
\begin{align}\label{Schrödinger R 2}
    i\frac{\partial\Psi^R_{k}[\phi_{k}^R,\phi_{k}^{R*},\tau]}{\partial\tau}=\left(-\frac{\partial^2}{\partial\phi^{R *}_{k}\partial\phi^R_{k}}+k^2 |\phi^R_{k}|^2 \right)\Psi^R_{k}[\phi_{k}^R,\phi_{k}^{R*},\tau]
\end{align}
that admits the following ground state solution
\begin{align}
    (\Psi^R_{k})_0[\phi_{k},\phi_{k}^{*},\tau]=\mathcal{N}_k\exp\left(-k\phi^{R *}_{k} \phi^R_{k}-i{k}\tau\right).\label{psiR}
\end{align}
As expected, the vacuums defined by equations \eqref{psiM} and \eqref{psiR} at $\tau=0$ are essentially different. Nevertheless, both results are approximately equal when $\pi k/2a \gg 1$, which is equivalent to small accelerations. In this limit, the conformal factor is almost 1, so there is little difference between Rindler and Minkowski narratives. 

The temporal evolution of the vacuum \eqref{psiM} on a Cauchy hypersurface defined in the accelerated frame can be obtained considering the following ansatz for the ground state 
\begin{align}\label{param ground state M}
    (\Psi_{k}^M)_0[\phi_{k}^R,\phi_{k}^{R*},\tau]=N_{k}\exp\left(-k f_{k}(\tau)\phi_{k}^{R}\phi_{k}^{R*}+\Omega_{k}(\tau) \right),
\end{align}
with $\Omega_{k}(\tau)$ being an additive complex phase. As an initial condition, we impose that $f_{k}(0)=\coth\left(\frac{\pi k}{2a}\right)$ and $\Omega_{k}(0)=0$. The Schrödinger equation \eqref{Schrödinger R 2} for $(\Psi_{k}^M)_0$ gives us two equations, one for $f_{k}(\tau)$ and another for $\Omega_{k}(\tau)$. With the above initial conditions, the solutions are 
\begin{align}\label{f}
    f_{k}(\tau)=\coth\left(\frac{\pi k}{2a}+ik\tau\right)
\end{align} 
and 
\begin{align} \label{Omega}
    \Omega_{k}(\tau)=-\ln{\left[ \sinh{\left(\frac{\pi k}{2a}+ik\tau \right)}\right]},
\end{align}
where an integration constant in (\ref{Omega}) can be absorbed in the normalization factor.

\subsection{The de Broglie-Bohm approach}\label{dbb}
 
In this subsection, we describe the features of the quantum scalar field in the ground state \eqref{param ground state M} from the de Broglie-Bohm perspective. In the relativistic version of the Bohmian mechanics, the wave functional determines the time evolution of the Bohmian fields, which are not operators but actual fields evolving in spacetime, through the so-called guidance equations. The set of initial configurations for determining the field evolution is given by the squared norm of the wave functional at this initial time. Detailed analysis of the dBB theory in the context of Quantum Field Theory can be seen in \cite{Holland,Duerr,Struyve,Pinto-Neto:2013toa}. 

In order to obtain the Bohmian fields, we rewrite the wave functional (\ref{param ground state M}) in the polar form 
\begin{equation}
    \Psi_{k}=R_{k}e^{i S_{k}},\label{polar form}
\end{equation}
where $R_k$ and $S_k$ are the radial part and the phase, respectively. The wave functional (\ref{param ground state M}), after normalization, becomes  
\begin{align}\label{Psi radial}
    \Psi_k[\phi_k^R,\phi_k^{R*},\tau]=&\sqrt{\frac{k\Re[f_k(\tau)]}{\pi}}\exp\left\{-k\Re [f_k(\tau)]|\phi_k^R|^2+i\left(-k\Im[f_k(\tau)]|\phi_k^R|^2+\Im[\Omega_k(\tau)]\right)\right\},
\end{align}
where $\Re[f_k]$ and $\Im[f_k]$ are the real and imaginary parts of (\ref{f}), that is,
\begin{align}\label{Re fk}
    \Re[f_k(\tau)]=\frac{\sinh \left(\frac{\pi  k}{a}\right)}{\cosh \left(\frac{\pi  k}{a}\right)-\cos (2 k \tau)},\hspace{6.0mm}
    \Im[f_k(\tau)]=\frac{-\sin (2 k \tau)}{\cosh \left(\frac{\pi  k}{a}\right)-\cos (2 k \tau)},
\end{align}
and the the real and imaginary part of $\Omega_k(\tau)$ reads,
\begin{align}\label{OmegaReIm}
     \Re[\Omega_k(\tau)]=-\frac{1}{2}\ln\left[\cosh^2\left(\frac{\pi k}{2a}\right)-\cos^2(k\tau)\right],\hspace{2.0mm}
    \Im[\Omega_k(\tau)]=-\tan^{-1}\left(\coth\left(\frac{\pi k}{2a}\right)\tan(k\tau)\right).
\end{align}
Then $R_k$ and $S_k$ can be identified as
\begin{align}
     R_k(\phi_k^R,\phi_k^{R*},\tau)&=\sqrt{\frac{k\Re[f_k(\tau)]}{\pi}}\exp\left(-k\Re[f_k(\tau)]|\phi_k^R|^2\right),\label{Rk}\\
     S_k(\phi_k^R,\phi_k^{R*},\tau)&=-k\Im[f_k(\tau)]|\phi_k^R|^2+\Im[\Omega_k(\tau)].\label{Sk}
\end{align}
The Schrödinger equation \eqref{Schrödinger R 2} in terms of the wave function \eqref{polar form} yields two real equations, specifically
\begin{align}\label{real schrod}
  \frac{\partial S_k}{\partial \tau}+\frac{\partial S_k}{\partial\phi_k^{R*}}\frac{\partial S_k}{\partial\phi_k^R}+k^2|\phi_k^R|^2-\frac{1}{R_k}\frac{\partial^2R_k}{\partial\phi_{k}^{R}\partial\phi_{k}^{R*}}=0,\\
  \label{imag schrod}
    \frac{\partial R_{k}^2}{\partial\tau}+\frac{\partial}{\partial\phi_k^R}\left(R_{k}^2\frac{\partial S_k}{\partial\phi_k^{R*}}\right)+\frac{\partial}{\partial\phi_k^{R*}}\left(R_{k}^2\frac{\partial S_k}{\partial\phi_k^R}\right)=0.
    \end{align}

In the de Broglie-Bohm (dBB) quantum theory, the modes $\phi_{k}^{R}$ are assumed to be actual modes of the scalar field evolving in spacetime guided by the wave function $\Psi_k$ through the dBB guidance equations
\begin{align}
    \frac{\partial\phi_k^R}{\partial\tau}=\frac{\partial S_k}{\partial\phi_{k}^{R*}}&=-k\Im[f_k(\tau)]\phi_k^R\label{guidance eq 0}.
\end{align} 
These are first-order differential equations that give the time evolution in terms of the Rindler variable $\tau$, with one integration constant per mode given by some initial condition, which is not known and practically impossible to be determined experimentally (they are the hidden variables of the theory). However, assuming that at some initial time $t_0$ the probability density distribution of initial conditions is given by the Born rule, $P(\phi_{k}^{R}(t_0))=R^{2}_{k}(\phi_{k}^{R},t_0)$, then Eq.~\eqref{imag schrod} together with the guidance equations \eqref{guidance eq 0} guarantee that $R^{2}_{k}(\phi_{k}^{R},t)$ gives the probability density that the field mode has the value $\phi_{k}^{R}$ at time $t$. In this way, all statistical predictions of quantum theory can be recovered. Equation \eqref{imag schrod} can then be understood as a continuity equation for an ensemble of field trajectories in configuration space with probability distribution $P=R_k^2$ and velocity field given in Eq.~\eqref{guidance eq 0}.

Once the guidance equations \eqref{guidance eq 0} are settled down, one can read Eq.~\eqref{real schrod} as a Hamilton-Jacobi equation for the mode dynamics, supplemented by an extra term. When this extra term becomes negligible with respect to the others, the classical evolution is recovered. From Hamilton-Jacobi theory, the energy is associated with $E_k=-\frac{\partial S_k}{\partial t}$. Considering that the modes with wave numbers $k$ and $-k$ have the same contribution and were counted twice in \eqref{Hamiltonian R 2}, the effective contribution to the energy in the Hamilton-Jacobi equation for each wave number should be divided by 2. Hence, looking at Eq.~\eqref{real schrod}, we define:

\begin{align}
    E_k(\tau)\equiv-\frac{1}{2}\biggl{(}\frac{\partial S_k}{\partial t}\biggl{)}=\frac{1}{2}\biggl{(}k\frac{\partial \Im[f_k(\tau)]}{\partial \tau}|\phi_k^R|^2-\frac{\partial \Im[\Omega_k(\tau)]}{\partial \tau}\biggl{)},\label{Ek}
\end{align}

\begin{align}
    K_k(\tau)\equiv \frac{1}{2}\biggl{(}\frac{\partial S_k}{\partial\phi_k^{R*}}\frac{\partial S_k}{\partial\phi_k^R}\biggl{)}=\frac{1}{2}\biggl{(}k^2\Im^2[f_k(\tau)]|\phi_k^R|^2\biggl{)},\label{kin}
\end{align}

\begin{align}
    V_k(\tau)\equiv\frac{1}{2}\biggl{(}k^2|\phi_k^R|^2\biggl{)}, \label{Cpot RighRindWedg}
\end{align}

\begin{align}\label{Qpot RighRindWedg}
    Q_k(\tau)\equiv\frac{1}{2}\biggl{(}-\frac{1}{R_k}\frac{\partial^2R_k}{\partial\phi_{k}^{R}\partial\phi_{k}^{R*}}\biggl{)}=\frac{1}{2}\biggl{(}k\Re[f_k(\tau)]-k^2\Re^2[f_k(\tau)]|\phi_k^R|^2\biggl{)},
\end{align}
with

\begin{align}\label{total energy}
    E_k(\tau)=K_k(\tau)+V_k(\tau)+Q_k(\tau).
\end{align}
From Eq.~\eqref{real schrod}, Eq. \eqref{Ek} gives the total energy of field mode, Eq.~\eqref{kin} can be viewed as the "classical" kinetic term, Eq.~\eqref{Cpot RighRindWedg} is the "classical" potential term, and Eq.~\eqref{Qpot RighRindWedg} refers to the so-called quantum potential. Their complete expressions are given in Appendix A, together with the asymptotic expansions for high and low accelerations of the coefficients that appear in the wave functional \eqref{Psi radial}. As mentioned above, when the quantum potential term is negligible with respect to the others, the classical evolution is recovered. 

Note that, for $a<<1$ and using Eqs.~(\ref{Re fk},\ref{OmegaReIm}), we recover the expressions for the total energy of the Minkowski vacuum and its parts in the dBB approach: $E_k=k/2$, $K_k\approx 0$, and $Q_k=k/2 - V_k$.

An important remark: this separation of terms in the total energy makes sense only in the dBB approach for quantum theory.

Computing the derivative of the Hamilton-Jacobi equation \eqref{real schrod} with respect to $\phi_k^{R*}$ and using the guidance equation, we obtain a Klein-Gordon type equation for the Bohmian field, with a source term due to the quantum potential, that is,
\begin{align}
    \frac{\partial^2 \phi_k^R}{\partial \tau^2} + k^2\phi_k^R&=-2\frac{\partial Q_k}{\partial \phi_k^{R*}},\label{KG bohmian}
\end{align}
introducing to the Bohmian field dynamics a corrective possibly non-linear quantum force.

\subsection{Mean values and Unruh temperature}\label{Mean values}

In the previous subsection, we saw that $R^2=|\Psi|^2$ in the dBB quantum theory is interpreted as the probability density associated with an ensemble of trajectories given in terms of the guidance equations. Hence, 
\begin{align}
    \braket{\mathcal{O}(t)}_{dBB}=\int d\phi_k^Rd\phi_k^{R*}\left|\Psi^R_{k}[\phi_{k}^R,\phi_{k}^{R*},\tau]\right|^2\mathcal{O}(\phi_{k}^R,\phi_{k}^{R*},\tau)\label{mean}
\end{align}
is the mean value of a physically meaningful property $\mathcal{O}$ related to the field trajectories \cite{Holland}. In order to give the same results as the usual mean values of quantum operators, the property $\mathcal{O}$, which is not an operator, must be judiciously chosen. For instance, in the present case, mean values of the Hamiltonian operator ${\hat{H}}_k$ are proven to be equal to the mean value of the property $\mathcal{O}=E_k$ as defined in Eq.~\eqref{Ek}. However, as commented above, the present formalism allows the differentiation of the classical parts from contributions of a quantum nature which the ensemble average might have (in this case, given by the quantum potential), which is not possible in the usual formalism. Hence, the mean energy $\braket{E_k}_{dBB}$ can be written as
\begin{align}\label{HJ aver sum}
    \braket{E_k}_{dBB}=\braket{K_k}_{dBB}+\braket{V_k}_{dBB}+\braket{Q_k}_{dBB},
\end{align}
where we have an explicit term due to the quantum potential.

Using Eq.~\eqref{mean} with $\mathcal{O}=E_k$ defined in Eq.~\eqref{Ek} yields the mean energy as
\begin{align}\label{meanEnergy}
    \braket{E_k}_{dBB}=\frac{k}{2}\coth\left(\frac{\pi k}{a}\right)=k\left(\frac{1}{2}+\frac{1}{e^{\frac{2\pi}{a}k}-1}\right),
\end{align}
where we used the fact that $\mathop{\mathlarger{\int}}_0^{\infty}d\rho \rho^3e^{-c\rho^2}=\dfrac{1}{2c^2}$, with $\rho=\left|\phi_k^R\right|$ and $c=2k\Re[f_k(\tau)]$.

We can explore this result to obtain the mean number of Rindler particles in the Minkowski vacuum if we use the known fact that ${\hat{H}}_k=\left({\hat{n}}_k+\frac{1}{2}\right)k$. Taking the average on both sides we have that $\braket{n_k}_{dBB}=\frac{1}{k}\braket{E_k}_{dBB}-\frac{1}{2}$, yielding, 
\begin{align}\label{nk}
    \braket{n_k}_{dBB}=\frac{1}{e^{\frac{2\pi}{a}k}-1},
\end{align}
which is the Bose-Einstein distribution with Unruh temperature $T=a/2\pi$.

For very low temperatures (accelerations), the exponential in the denominator of Eq.~\eqref{nk} goes to infinity, so that the mean occupation number of Rindler particles in the Minkowski vacuum is null, meaning that the Rindler and Minkowski vacua are equivalent. Consequently, the energy average is just the energy of a single harmonic oscillator with wave number $k=\hbar w$ in the ground state, $\braket{E_k}_{dBB}=k/2$. For high temperatures (accelerations), in contrast, $\braket{E_k}_{dBB}=T=a/2\pi$, the average energy of a thermal distribution of oscillators at temperature $T$, agreeing with the equipartition theorem. 

Let us now calculate the mean values of the different parts of the energy from Eqs.~(\ref{kin},\ref{Cpot RighRindWedg},\ref{Qpot RighRindWedg}). They read:

\begin{align}\label{Kkmean} 
\braket{K_k}_{dBB}&=\frac{k}{4}\frac{\Im^2[f_k(\tau)]}{\Re{[f_k(\tau)]}}=\frac{k \csch\left(\frac{\pi  k}{a}\right) \sin ^2(2 k t)}{4 \left[\cosh \left(\frac{\pi  k}{a}\right)-\cos (2 k t)\right]}, \\
\label{Vkmean}\braket{V_k}_{dBB}&=\frac{k}{4}\frac{1}{\Re{[f_k(\tau)]}}=\frac{k \left[\cosh \left(\frac{\pi  k}{a}\right)-\cos (2 k t)\right]}{4 \sinh \left(\frac{\pi  k}{a}\right)},
\end{align}
and
\begin{align}\label{Qkmean}
\braket{Q_k}_{dBB}=\frac{k}{4}\Re{[f_k(\tau)]}=\frac{k \sinh \left(\frac{\pi  k}{a}\right)}{4 \left[\cosh \left(\frac{\pi  k}{a}\right)-\cos (2 k t)\right]}.
\end{align}

\begin{figure}[ht]
     \centering
     \begin{subfigure}[b]{0.4\textwidth}
         \centering
         \includegraphics[width=0.96\textwidth]{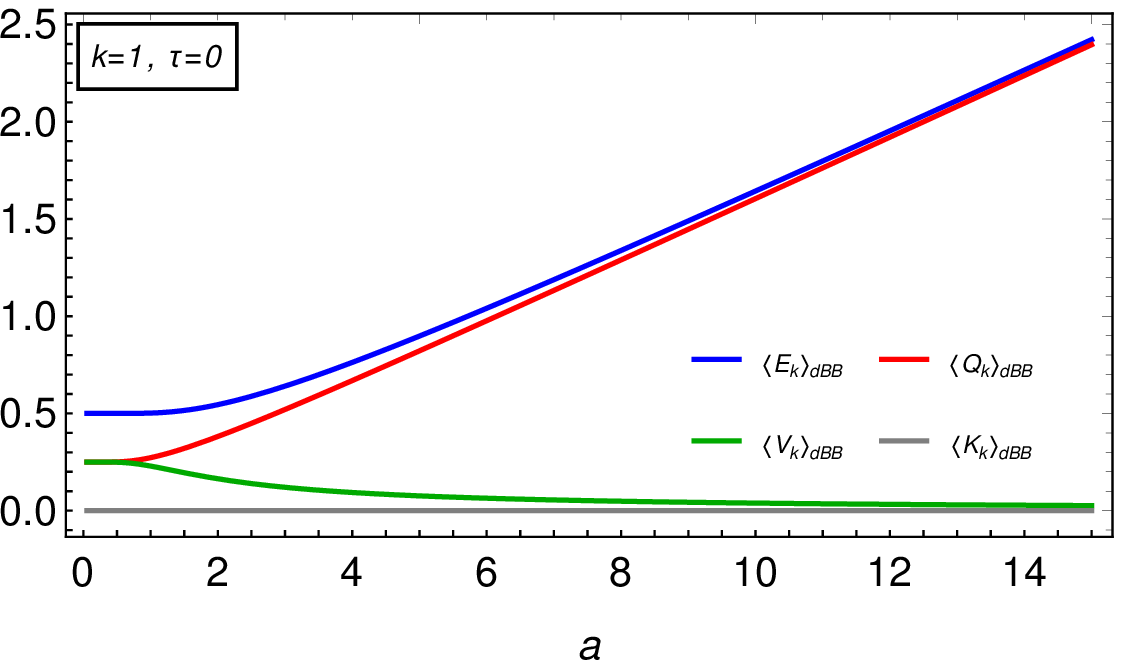}
         \caption{$k=1.0$ and $\tau=0$}
         \label{E_vs_a_t0}
     \end{subfigure}
     \centering
\begin{subfigure}[b]{0.4\textwidth}
         \centering\includegraphics[width=0.95\textwidth]{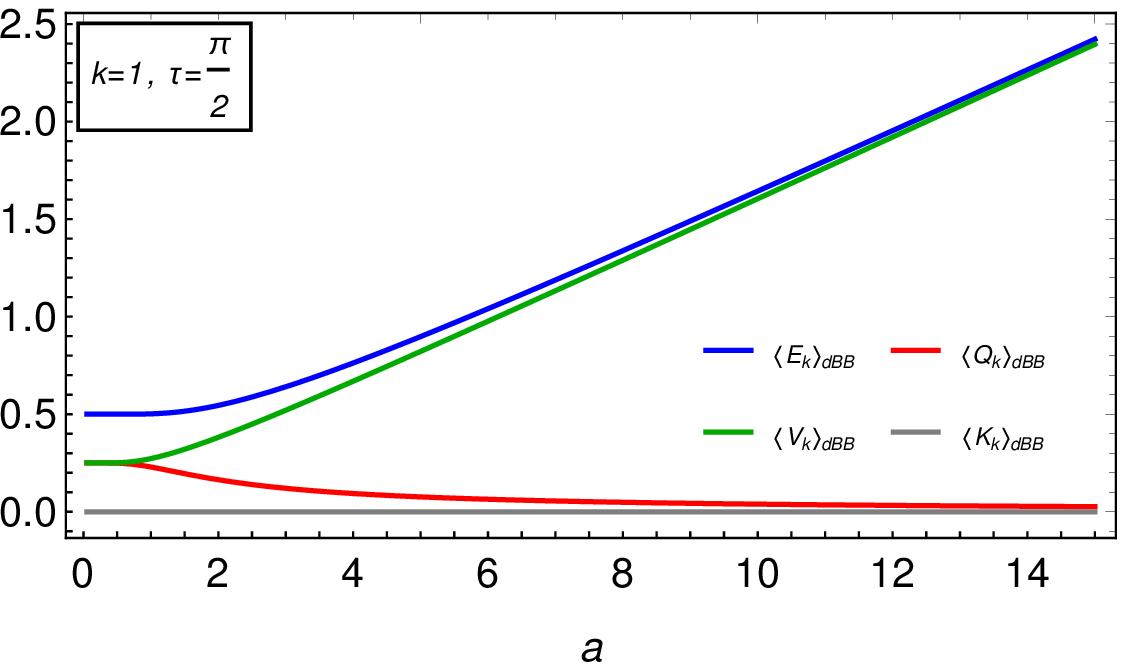}
         \caption{$k=1.0$ and $\tau=\pi/2$}
         \label{E_vs_a_t1}
\end{subfigure}
     \caption{The mean values as functions of the acceleration $a$ for (\ref{E_vs_a_t0}) $\tau=0$ and (\ref{E_vs_a_t1}) $\tau=\pi/2$.}
     \label{E_vs_a}
\end{figure}

Note that each of the above expressions has a non-trivial time dependence, but their sum, the total mean energy, is time-independent. In Figure (\ref{E_vs_a}) we plot the averages versus the acceleration $a$ for $\tau=0$ and $\tau=\pi/2$. Despite at the beginning classical and quantum potentials having almost the same value in both cases, for high accelerations the quantum potential tends asymptotically to $\braket{E_k}_{dBB}$ in Figure (\ref{E_vs_a_t0}), while in Figure (\ref{E_vs_a_t1}) the main responsible for the mean energy is the classical potential.

Let us see the limits of these averages for low and high temperatures:

\subsection*{Low temperature (acceleration) regime: $T<<1$}

In this regime we have:

\begin{eqnarray}
\braket{K_k}_{dBB}&\approx& k\,{\sin}^2(2k\tau)e^{-k/T}\approx 0,\\
\braket{V_k}_{dBB}&\approx& \frac{k}{4}-\frac{k\cos(2k\tau)e^{-k/(2T)}}{2} \approx \frac{k}{4},\\
\braket{Q_k}_{dBB}&\approx& \frac{k}{4}+\frac{k\cos(2k\tau)e^{-k/(2T)}}{2} \approx \frac{k}{4}.
\end{eqnarray}
In this case we recover the usual dBB picture of the vacuum state: the energy of the field is equally shared between the classical and quantum potential, with negligible kinetic energy.

\subsection*{High temperature (acceleration) regime: $T>>1$}

In this case, we have two different situations:
\vspace{0.5cm}
\subsection*{i) $\boldsymbol{\tau\neq n\pi/k}$, with $\boldsymbol{n}$ an integer}
\vspace{0.5cm}

The results are:

\begin{eqnarray}
\braket{K_k}_{dBB}&\approx& T\,{\cos}^2(k\tau)\\
\braket{V_k}_{dBB}&\approx& T\,{\sin}^2(k\tau)\\
\braket{Q_k}_{dBB}&\approx& \frac{k^2}{16 T {\sin}^2(k\tau)}\approx 0
\end{eqnarray}
Note that in this limit the classical kinetic and potential energies supply all the total energy $T$, with a negligible contribution of the quantum potential.

\vspace{0.5cm}
\subsection*{ii) $\boldsymbol{\tau = n\pi/k}$, with $\boldsymbol{n}$ an integer}
\vspace{0.5cm}

The situation changes drastically when $\tau = n\pi/k$, with $n$ an integer. In this case we have:
\begin{eqnarray}
\braket{K_k}_{dBB}&\approx& 0\\
\braket{V_k}_{dBB}&\approx& \frac{k^2}{16 T}\approx 0\\
\braket{Q_k}_{dBB}&\approx&  T 
\end{eqnarray}
At these times, the quantum potential is the main contribution to the total energy, while the classical terms yield negligible contributions. 
Hence, the total mean energy is constant, but there is a significant shift from classical to quantum contribution occurring periodically at $\tau = n\pi/k$, which instigates us to think about the possibility of measuring such an effect.

As a matter of fact, these abrupt changes can be seen already in the effective Klein-Gordon equation for the Bohmian field
\begin{align}
    \frac{\partial^2 \phi_k^R}{\partial \tau^2} + k^2\phi_k^R&=-2\frac{\partial Q_k}{\partial \phi_k^{R*}}=k^2\Re^2[f_k(\tau)]\phi_k^R.\label{KG bohmian2}
\end{align}
As indicated by Eq.~\eqref{KG bohmian2}, in the case of the wave function \eqref{Psi radial}, the source term is linear, playing the role of an effective mass. In the high temperature regime, $T>>1$, and for 
$\tau\neq n\pi/k$, we get:

\begin{align}\label{quantum force1}
    \frac{\partial^2 \phi_k^R}{\partial \tau^2} + k^2\phi_k^R
    \approx \frac{k^4}{16 T^2 {\sin}^2(k\tau)}\phi_k^R\approx 0\; ;\;\;\;\;\tau\neq n\pi/k.
\end{align}
The quantum force is negligible, and the Bohmian field obeys a classical Klein-Gordon equation. However, for $\tau = n\pi/k$, one gets that

\begin{align}
    \frac{\partial^2 \phi_k^R}{\partial \tau^2} + k^2\phi_k^R \approx 16T^{2}\phi_k^R\; ;\;\;\;\;\tau = n\pi/k\label{quantum force2},
\end{align}
and now the quantum force drives the field dynamics. Hence, also in this perspective, there is a substantial change from classical to quantum dominance in the neighborhood $\tau = n\pi/k$. Since the field dynamics are different for these two distinct moments, it allows us to speculate whether such an effect can be observed.

\begin{figure}[ht]
     \centering
     \includegraphics[width=0.5\textwidth]{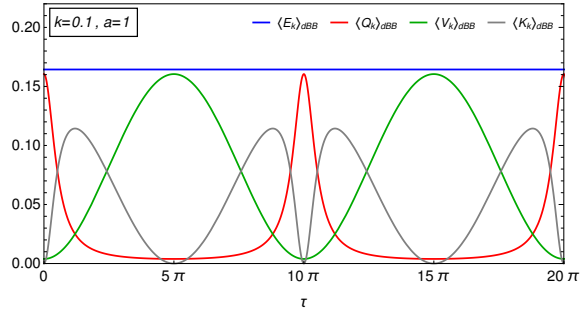}
     \caption{Plot of the average values for $k=0.1$ and $a=1$. Despite the dominance of the quantum potential around $\tau=n\pi/k$, the classical terms are still relevant in this case, with a non-negligible contribution.}
     \label{fig: a1}
\end{figure}

\begin{figure}[ht]
     \centering
     \includegraphics[width=0.55\textwidth]{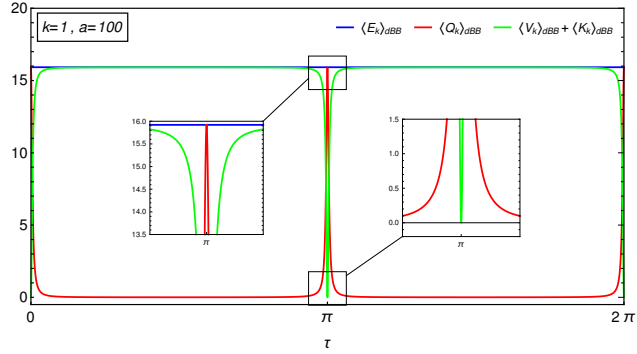}
     \caption{Quantum and classical averages, together with the total mean energy $\braket{E_k}_{dBB}$, for $k=1$ and $a=10^2$. In the vicinity of $\tau=n\pi/k$, there is an abrupt change in the quantum and classical contributions, with $\braket{Q_k}_{dBB}$ rapidly becoming the dominant part of $\braket{E_k}_{dBB}$, while $\braket{V_k}_{dBB}+\braket{K_k}_{dBB}$ suddenly dropping to zero in a very short range of time.}
     \label{a100}
\end{figure}

In Figure (\ref{fig: a1}) we plot all mean energies for accelerations of order $1$. In Figure (\ref{a100}), on the other hand, we plot the sum of the classical energies together with the quantum potential in the case of high accelerations (temperatures): $a=10^2$. One can see the periodic abrupt jumps from classical to quantum dominance in the neighborhoods of $\tau = n\pi/k$.

\subsection{Field trajectories} \label{field trajectories rhw}
In this subsection, we obtain the general solution of the guidance equations in order to calculate the ensemble of possible field trajectories. We show that there is a very particular one with remarkable properties, and we analyze its behavior within the limits of low and high acceleration. 

Integrating the guidance equation \eqref{guidance eq 0}, we obtain an explicit expression for the Bohmian field,
\begin{eqnarray}
    \phi_k^R(\tau)=C_k(a)\sqrt{\cosh\left(\frac{\pi k}{a}\right)-\cos(2k\tau)}.
\end{eqnarray}
Without loss of generality, we write the integration constant as $C_k(a)=D_k(a)/[2k\sinh((\pi k/a))]^{1/2}$. Therefore, the field trajectory reads\footnote{Strictly speaking, we should use the absolute value of the wave number in the exponents due to the reality of the field, as can be seen by dimensional analysis. However, admitting decomposition \eqref{psi decomposition}, we are restricted to positive values of $k$, making the module's presence unnecessary.},
\begin{align}\label{bohmian field}
    \phi_k^R(\tau)=\frac{D_k(a)}{\sqrt{2k\Re[f_k(\tau)]}}=D_k(a)\sqrt{\frac{\cosh\left(\frac{\pi k}{a}\right)-\cos(2k\tau)}{2k\sinh\left(\frac{\pi k}{a}\right)}}.
\end{align}

Written in these terms, the probability density distribution admits a straightforward form, the Gaussian $|\Psi_k|^2\propto e^{-|D_k(a)|^2}$. The total energy of this ensemble of fields reads

\begin{align}\label{EkD}
E_k=k\frac{|D_k(a)|^2-1 + \cosh \left(\frac{\pi  k}{a}\right) 
\left[\cosh \left(\frac{\pi  k}{a}\right)-|D_k(a)|^2 \cos (2 k \tau)\right]}{2 \left[\cosh \left(\frac{\pi  k}{a}\right)-\cos (2 k \tau)\right] \sinh \left(\frac{\pi  k}{a}\right)}.
\end{align}

From Eq.~\eqref{EkD}, one can immediately see that the total energy of the Bohmian fields is time independent if and only if we make the choice $|D_k(a)|^2=1$, so that $D_k(a)$ must be just a phase, $D_k(a)=\exp(i \theta_k(a))$. In this case, disregarding the normalization factor, the Gaussian distribution part $|\Psi_k|^2\propto e^{-|D_k(a)|^2}$ of initial conditions for this subset of possibilities is fixed and it is independent of $k$ and $a$. The energy of this particular subset emerging from Eq.~\eqref{EkD} is precisely the mean energy given in \eqref{meanEnergy}:
\begin{align}
    E_k=k\left(\frac{1}{2}+\frac{1}{e^{\frac{2\pi}{a}k}-1}\right).
\end{align}
Furthermore, all components of the total energy of this particular Bohmian field are exactly the average values calculated in the last section, that is, $Q_k=\braket{Q_k}_{dBB}$, $V_k=\braket{V_k}_{dBB}$ and $K_k=\braket{K_k}_{dBB}$ (equations \eqref{Qkmean}, \eqref{Vkmean} and \eqref{Kkmean}). Hence, the analysis corresponding to the asymptotic limits of the average quantities made in the previous subsection is also valid for every single Bohmian field with $D_k(a)=\exp(i \theta_k(a))$, including the periodic abrupt transitions from classical to quantum dominance discussed above. These particular Bohmian fields follow the mean value evolution exactly.

Finally, the asymptotic behaviors of these particular Bohmian fields, disregarding their phase, read

\begin{align}\label{Bohmian field} 
\phi_k^R&=\frac{1-\cos(2 k \tau) e^{-k/(2T)}}{\sqrt{2k}}\; , \;\;\;\;\; T<<1\\
\phi_k^R&=\frac{\sqrt{2T}|\sin(2 k \tau)|}{k}\; , \;\;\;\;\; T>>1
\end{align}

\subsection{Power Spectrum}\label{power spectrum rhw}

In order to present statistical predictions, we calculate the two-point correlation function, then integrate it over the phase space. We denote $\phi(\tau,\xi;\phi_i)$ as a solution to the guidance equations with the initial condition $\phi_i$. In dBB interpretation, the two-point correlation function is calculated as an average over all initial field configurations with the weight $|\Psi(\phi_i,\tau_i)|^2$. If the initial field $\phi_i$ is distributed according to quantum equilibrium $|\Psi(\phi_i,\tau_i)|^2$ then $\phi(\tau,\xi;\phi_i)$ is distributed according to $|\Psi(\phi,\tau)|^2$ at any time \cite{Durr-Teufel, Colin:2014pna}, and it is possible to show that \cite{Pinto-Neto:2013npa}
\begin{align}\nonumber
    \left<\phi(\tau,\xi)\phi(\tau,\xi+\sigma)\right>_{dBB} & =\int D\phi_i |\Psi(\phi_i(\tau_i,\xi))|^2\phi(
    \tau,\xi;\phi_i)\phi(\tau,\xi+\sigma;\phi_i)\\
    \label{two-p f 1}& =\int D\phi |\Psi(\phi(\tau,\xi))|^2\phi(\xi)\phi(\xi+\sigma). 
\end{align}
This means that the two-point function in dBB interpretation is the same as the one calculated in the usual manner. In the case of the general parameterization of the ground state \eqref{param ground state M}, the integral \eqref{two-p f 1} results in
\begin{align}\label{two-p f 2}
     \left<\phi(\xi)\phi(\xi+\sigma)\right>_{dBB}=\frac{1}{2\pi}\int_{-\infty}^{\infty} dk\frac{1}{2|k| \Re[f_k(\tau)]} e^{-i k \sigma}.
\end{align}

Using the expression (\ref{two-p f 2}), the power spectrum 
\begin{align}\label{P}
    P_k(\tau)=\int d\xi e^{-ik\xi}\left<\phi(\xi)\phi(0)\right>_{dBB},
\end{align}
for $\xi +\sigma=0$ is equal to
\begin{align}\label{Power spectrum}
    P_k(\tau)=\frac{1}{2k \Re[f_k(\tau)]}=\frac{\cosh\left(\frac{\pi k}{a}\right)-\cos(2 k \tau)}{2k\sinh\left(\frac{\pi k}{a}\right)}=\frac{2}{k^2}\braket{V_k}_{dBB}.
\end{align}
From the last equality, one can see that the correlations between the field modes are closely connected to the classical potential (see Eq.~\eqref{Vkmean}).

In the high temperature regime $P_k(\tau)$ can be approximated by
\begin{align}
    P_k(\tau)\simeq\frac{2T}{k^2}\sin^2(k\tau),
\end{align}
being independent of time for low temperatures because, in this limit,
\begin{align}
    P_k(\tau)\simeq\frac{1}{2k}.
\end{align}

\section{Complete manifold analysis} \label{complete analysis}
In this section, we analyze the complete manifold problem, i.e. the two-wedge scalar massless field in $(1+1)-$dimensions, giving de Broglie-Bohm's prescription of the Unruh effect. The two-wedge approach introduces new properties to the Unruh effect, which appears in the Hawking radiation, since a non-local connection between the fields defined in the two wedges \cite{Unruh-Wald}, as the two-wedge Rindler geometry gets features of a Schwarzschild-like geometry. Since the dBB quantum theory is manifestly non-local, this alternative approach offers direct regard to these new features, which may be useful for future analysis.

Let us consider the expansion of the Minkowski field in plane waves
\begin{align}
\phi(t,x)=\int_{-\infty}^{\infty}\frac{dk}{\sqrt{2\pi}}e^{ikx}\phi_{k}^M(\tau)\label{FourierM2}.
\end{align}
Using the decomposition \eqref{psi decomposition}, the associated ground state wave functional becomes 
\begin{align}
(\Psi_{k}^M)_0[\phi_k^M,t]=N_k\exp\left(-k\left(\phi^{M}_k\right)^2-ikt \right).\label{ground state M complete2}
\end{align}

With the purpose of describing the entire Minkowski space, we analytically extend the right Rindler wedge to the left side by introducing the two-wedge coordinates:
\begingroup
\setlength{\tabcolsep}{10pt}
\renewcommand{\arraystretch}{2.2} 
\begin{align}
\begin{tabular}{ l l }
RH-wedge ($x>0$): & LH-wedge ($x<0$): \\
$x=\dfrac{e^{a \xi_R}}{a}\cosh{(a\tau)}$ & $x=-\dfrac{e^{a \xi_L}}{a}\cosh{(a\tau)}$\\
$t=\dfrac{e^{a \xi_R}}{a}\sinh{(a\tau)}$ & $t=-\dfrac{e^{a \xi_L}}{a}\sinh{(a\tau)}.$\\
\end{tabular}
\end{align}
\endgroup
In the LH-wedge, the time parameter $\tau$ evolves in the opposite direction, therefore, it can be considered a time-reversed copy of the RH-wedge \cite{Valdes,Matsas}. The field expansion is Rindler variables is
\begin{align}
\phi(\tau,\xi)=\theta(x)\int_{-\infty}^{\infty}\frac{dk}{\sqrt{2\pi}}e^{ik\xi_R}\phi_{k}^R(\tau)+\theta(-x)\int_{-\infty}^{\infty}\frac{dk}{\sqrt{2\pi}}e^{ik\xi_L}\phi_{k}^L(\tau),\label{FourierR expanded}
\end{align}
with $\phi_{k}^R$ equal to the right modes as in the previous case, and $\phi_{k}^L$ corresponding to the left modes. Here, $\theta(x)$ is the step function.

Proceeding similarly as in section \ref{KG}, we obtain the following wave functional at $t=0$ \cite{Freese}
\begin{align}
  \left(\Psi_k(\tau=t=0)\right)_0=N_k\exp\left[-k \coth\left(\frac{\pi k}{a}\right)\left(|\phi_{k}^R|^2+|\phi_{k}^L|^2\right)+k\csch\left(\frac{\pi k}{a}\right)\left(\phi_{k}^R\phi_{k}^{L*}+\phi_{k}^L\phi_{k}^{R*}\right)\right], \label{Rindler wave fnctional extended tau=0}
\end{align}
with $N_k$ a normalization constant. Since at $t=\tau=0$ both Rindler and Minkowski spaces share the same Cauchy hypersurface, we can use equation \eqref{Rindler wave fnctional extended tau=0} as an initial condition for the Schrödinger equation
\begin{align}\label{schrod eq ext}
    i\frac{\partial \Psi_{k}(\tau)}{\partial \tau}=\left(-\frac{\partial^2}{\partial\phi_k^{R*}\partial\phi_k^{R}}-\frac{\partial^2}{\partial\phi_k^{L*}\partial\phi_k^{L}}+k^2\left(|\phi_k^R|^2+|\phi_k^L|^2\right) \right)\Psi_{k}(\tau).
\end{align}
In this case, the general solution is
\begin{align}\label{vac solution}
    \left(\Psi_k(\tau)\right)_0 = N_k\exp\left(-k F_k(\tau) \left(|\phi_k^R|^2+|\phi_k^L|^2\right) +k G_k(\tau)\left(\phi_k^{R*}\phi_k^{L}+\phi_k^{R}\phi_k^{L*} \right) +\Theta_k(\tau) \right), 
\end{align}
with the coefficients 
\begin{align}\label{coef}
    F_k(\tau)&=\coth{\left(\frac{\pi k}{a}+2ik\tau\right)},\hspace{6.0mm}G_k(\tau)=\csch{\left(\frac{\pi k}{a}+2ik\tau\right)},
\end{align}
the phase 
\begin{align}\label{phase}
    \Theta_k(\tau)=-\ln{\left[ \sinh{\left(\frac{\pi k}{a}+2ik\tau \right)}\right]},
\end{align}
and the normalization constant
\begin{align}\label{norm}
N_k(\tau)=\frac{k}{\pi}\sinh\left(\frac{\pi  k}{a}\right).
\end{align}
To our knowledge, this is the first time Eq.~\eqref{vac solution}, together with Eqs.~(\ref{coef},\ref{phase},\ref{norm}), is exhibited, which is the mode $k$ Minkowski vacuum wave function solution in terms of the Rindler fields in both wedges for any Rindler time $\tau$. 

The presence of the crossed terms in the wave functional \eqref{vac solution} indicates the existence of a non-trivial correlation between $\phi_k^R$ and $\phi_k^L$ \cite{Unruh-Wald}. Despite the horizon at $t=\pm x$, the right-wedge field depends on the left-wedge field, even without interaction between them. This mutual dependence can be understood when we look at the dBB guidance equations. 

Turning to the Bohmian theory of the extended case and writing the wave functional in the polar form $\left(\Psi_k(\tau)\right)_0=R_ke^{iS_k}$, we obtain 
\begin{align}
     R_k&=N_k\exp\left(-k\Re[F_k(\tau)]\left(|\phi_k^R|^2+|\phi_k^L|^2\right)+k\Re[G_k(\tau)]\left(\phi_k^R\phi_k^{L*}+\phi_k^L\phi_k^{R*}\right)+\Re[\Theta_k(\tau)]\right),\\
     S_k&=-k\Im[F_k(\tau)]\left(|\phi_k^R|^2+|\phi_k^L|^2\right)+k\Im[G_k(\tau)]\left(\phi_k^R\phi_k^{L*}+\phi_k^L\phi_k^{R*}\right)+\Im[\Theta_k(\tau)],
\end{align}
where the calligraphic letters $\Re$ and $\Im$ refer to the real and imaginary parts of their respective coefficients, which are given by
\begin{align}\label{Re FFk}
    \Re[F_k(\tau)]&=\frac{\sinh \left(\frac{2\pi  k}{a}\right)}{\cosh \left(\frac{2\pi  k}{a}\right)-\cos (4 k \tau)},\hspace{8.0mm}
    \Im[F_k(\tau)]=\frac{-\sin (4 k \tau)}{\cosh \left(\frac{2\pi  k}{a}\right)-\cos (4 k \tau)},\\
    \Re[G_k(\tau)]&=\frac{2\sinh{\left(\frac{\pi k}{a}\right)}\cos{(2k\tau)}}{\cosh{\left(\frac{2\pi k}{a}\right)}-\cos(4k\tau)},\hspace{8.0mm}
    \Im[G_k(\tau)]=\frac{-2\cosh{\left(\frac{\pi k}{a}\right)}\sin{(2k\tau)}}{\cosh{\left(\frac{2\pi k}{a}\right)}-\cos(4k\tau)},\label{Re Gk}
\end{align}
while for the phase, we have that
\begin{align}\label{Re theta}
    \Re[\Theta_k(\tau)]&=-\frac{1}{2}\ln\left[\cosh^2\left(\frac{\pi k}{a}\right)-\cos^2(2k\tau)\right],\\
    \Im[\Theta_k(\tau)]&=-\tan^{-1}\left(\coth\left(\frac{\pi k}{a}\right)\tan(2k\tau)\right). 
\end{align}
Then, the Hamilton-Jacobi and continuity equations are, respectively, 
\begin{align}\label{real schrod ext}
  \frac{\partial S_k}{\partial \tau}+\sum_{a=R,L}\left[\left(\frac{\partial S_k}{\partial\phi_k^{a*}}\frac{\partial S_k}{\partial\phi_k^a}\right)+k^2|\phi_k^a|^2\right]
  -\frac{1}{R_k}\sum_{a=R,L}\left(\frac{\partial^2R_k}{\partial\phi_{k}^{a}\partial\phi_{k}^{a*}}\right)=0,\\
  \label{imag schrod ext}
    \frac{\partial R_{k}^2}{\partial\tau}+\sum_{a=R,L}\left[\frac{\partial}{\partial\phi_k^a}\left(R_{k}^2\frac{\partial S_k}{\partial\phi_k^{a*}}\right)+\frac{\partial}{\partial\phi_k^{a*}}\left(R_{k}^2\frac{\partial S_k}{\partial\phi_k^a}\right)\right]=0.
     \end{align}
Consequently, we interpret $R_k^2$ as a probability distribution and $\frac{\partial S_k}{\partial \phi_k^a}$, with $a=R,L$, as the velocity fields. Hence, the $dBB$ guidance equations are
\begin{align}\label{guidance eq 1 extended}
    \frac{\partial\phi_k^R}{\partial\tau}=\frac{\partial S_k}{\partial\phi_{k}^{R*}}=-k\Im[F_k(\tau)]\phi_k^R+k\Im[G_k(\tau)]\phi_k^L,\\
    \label{guidance eq 2 extended}
    \frac{\partial\phi_k^L}{\partial\tau}=\frac{\partial S_k}{\partial\phi_{k}^{L*}}=-k\Im[F_k(\tau)]\phi_k^L+k\Im[G_k(\tau)]\phi_k^R,
\end{align}
which reveals that the right and left modes have, at a first glance, a non-local connection. The change in $\phi_k^L$ has an immediate effect on $\phi_k^R$ through the guidance equations and vice-versa. This can be seen also from the effective Klein-Gordon equations for the Bohmian fields,
\begin{align}
\frac{\partial^2 \phi_k^R}{\partial \tau^2} + k^2\phi_k^R&=k^2(\Re^2[f_k]+\Re^2[g_k])\phi_k^R-2k^2\Re[f_k]\Re[g_k]\phi_k^L \label{force ext R}\\
\frac{\partial^2 \phi_k^L}{\partial \tau^2} + k^2\phi_k^L&=k^2(\Re^2[f_k]+\Re^2[g_k])\phi_k^L-2k^2\Re[f_k]\Re[g_k]\phi_k^R \label{force ext L}.
\end{align}

Nevertheless, this system can be decoupled by introducing the variables
\begin{align}
    \chi_{1,k}=\frac{\phi_k^R+\phi_k^L}{\sqrt{2}}, \hspace{6.0mm} \chi_{2,k}=\frac{\phi_k^R-\phi_k^L}{\sqrt{2}},\label{chi}
\end{align}
so that, in terms of $\chi_{k,1}$ and $\chi_{k,2}$ the guidance equations implies that
\begin{align}\label{guidance eq 1 extended xi}
    \frac{\partial\chi_{1,k}}{\partial\tau}=-k\Im[H_{1,k}(\tau)]\chi_{1,k},\\
    \label{guidance eq 2 extended xi}
    \frac{\partial\chi_{2,k}}{\partial\tau}=-k\Im[H_{2,k}(\tau)]\chi_{2,k},
\end{align}
where $H_{1,k}$ and $H_{2,k}$ are time-dependent coefficients with the following expressions
\begin{align}
    H_{1,k}&=F_k-G_k=\tanh{\left(\frac{\pi k}{2a}+ik\tau\right)},\\    H_{2,k}&=F_k+G_k=\coth{\left(\frac{\pi k}{2a}+ik\tau\right)}.
\end{align}
The corresponding real and imaginary parts of $H_{1,k}$ and $H_{2,k}$ are
\begin{align}\label{Re H1k}
    \Re[H_{1,k}(\tau)]&=\frac{\sinh \left(\frac{\pi  k}{a}\right)}{\cosh \left(\frac{\pi  k}{a}\right)+\cos (2 k t)},\hspace{6.0mm}
    \Im[H_{1,k}(\tau)]=\frac{\sin (2 k t)}{\cosh \left(\frac{\pi  k}{a}\right)+\cos (2 k t)},\\
    \Re[H_{2,k}(\tau)]&=\frac{\sinh \left(\frac{\pi  k}{a}\right)}{\cosh \left(\frac{\pi  k}{a}\right)-\cos (2 k t)},\hspace{6.0mm}
    \Im[H_{2,k}(\tau)]=-\frac{\sin (2 k t)}{\cosh \left(\frac{\pi  k}{a}\right)-\cos (2 k t)}.\label{Re H2k}
\end{align}

Thereafter, in terms of the new variables, the ground state \eqref{vac solution} can be written as the direct product between two independent states, that is,
\begin{align}\label{vac solution xi}
    \left(\Psi_k(\tau)\right)_0 &= \frac{k}{\pi}e^{\Theta_k(\tau)}\sinh{\left(\frac{\pi k}{a}\right)}e^{-k H_{1,k}(\tau) |\chi_{1,k}|^2}e^{-k H_{2,k}(\tau) |\chi_{2,k}|^2}\nonumber\\
    &\equiv\Psi_{1,k}[\chi_{1,k},\chi_{1,k}^*,\tau]\otimes\Psi_{2,k}[\chi_{2,k},\chi_{2,k}^*,\tau],
\end{align}
The wave functional $\Psi_{2,k}$ corresponds to a squeezed state \cite{Klaus} with a squeezing parameter $r_k$ such that $\tanh r_k=e^{-\pi k/a}$ and a squeezing angle $\alpha_k=-k\tau$, while $\Psi_{1,k}$ can be seen in the same way, but with the squeezing angle rotated by $\pi/2$. With this parameterization, we have 
\begin{align}\label{squeezed}
    \Psi_{A,k}[\chi_{A,k}]\propto\exp\left(-k\frac{1+e^{2i\alpha_k}\tanh r_k}{1-e^{2i\alpha_k}\tanh r_k}|\chi_{A,k}|^2\right),
\end{align}
with $A=1,2$.
Note that $H_{2,k}$ has the same expression as $f_k(\tau)$ in equation \eqref{f}. Then, $\Psi_{2,k}$ can be seen in the same manner as the right-wedge wave functional, but with $\phi_k^R$ substituted for $\chi_{2,k}$. Analogously, $\Psi_{1,k}$ is the left version of the ground state \eqref{param ground state M}. Therefore, the decomposition \eqref{vac solution xi} can be understood as the product of two decoupled Minkowski wave functionals for the mode $k$ in Rindler-like variables.

As in the previous section, one can write the total energy and its components in terms of the new field variables: 

\begin{align}\label{Ek ext}
    E_k(\tau)&=\frac{1}{2}\left[k\frac{\partial \Im[H_{1,k}(\tau)]}{\partial \tau}|\chi_{1,k}|^2+k\frac{\partial \Im[H_{2,k}(\tau)]}{\partial \tau}|\chi_{2,k}|^2-\frac{\partial \Im[\Theta_k(\tau)]}{\partial \tau}\right],
\end{align}

\begin{align}
    K_k(\tau)=\frac{1}{2}\left[k^2\Im^2[H_{1,k}(\tau)]|\chi_{1,k}|^2+k^2\Im^2[H_{2,k}(\tau)]|\chi_{2,k}|^2\right],\label{kin ext}
\end{align}

\begin{align}\label{Vk ext}
    V_k(\tau)&=\frac{1}{2}\left[k^2|\chi_{1,k}|^2+k^2|\chi_{2,k}|^2\right],
\end{align}

\begin{align}\label{Qk ext}
    Q_k(\tau)&=\frac{1}{2}\left[k\left(\Re[H_{1,k}(\tau)]+\Re[H_{2,k}(\tau)]\right)-k^2\left(\Re^2[H_{1,k}(\tau)]|\chi_{1,k}|^2+\Re^2[H_{2,k}(\tau)]|\chi_{2,k}|^2\right)\right],
\end{align}
which are, respectively, the total energy, the kinetic energy, the classical potential, and the quantum potential, from where we can see the individual contributions of $\chi_{1,k}$ and $\chi_{2,k}$.

Note that, for $a<<1$ and using Eqs.~(\ref{Re H1k},\ref{Re theta}), we recover the expressions for the total energy and its parts of two non-interacting fields in the Minkowski vacuum in the dBB approach: $E_k=k$, $K_k\approx 0$, and $Q_k=k - V_k$.

Lastly, the effective Klein-Gordon equations for the Bohmian fields, \eqref{force ext R} and \eqref{force ext L}, are decoupled in two independent equations, namely
\begin{align}
    \frac{\partial^2 \chi_{1,k}}{\partial \tau^2} + k^2\chi_{1,k}&=k^2\Re^2[H_{1,k}(\tau)]\chi_{1,k},\label{KG bohmian ext R}\\
    \frac{\partial^2 \chi_{2,k}}{\partial \tau^2} + k^2\chi_{2,k}&=k^2\Re^2[H_{2,k}(\tau)]\chi_{2,k},\label{KG bohmian ext L}
\end{align}
which, as before, are Klein-Gordon-type equations supplemented by a linear source of quantum origin.

\subsection{Mean values for the extended geometry}\label{Mean values ext}
The averages associated with the extended geometry field trajectories can be computed as
\begin{align}
    \braket{\mathcal{O}(t)}_{dBB}=\int D\phi_k\left|\Psi_{k}[\phi_{k}^R,\phi_{k}^{R*},\phi_{k}^L,\phi_{k}^{L*},\tau]\right|^2\mathcal{O}(\phi_{k}^R,\phi_{k}^{R*},\phi_{k}^L,\phi_{k}^{L*},\tau),
\end{align}
with $\mathcal{O}$ a meaningful property and $D\phi_k=d\phi_k^Rd\phi_k^{R*}d\phi_k^Ld\phi_k^{L*}$ is the modes integration measure. Since the wave functional \eqref{vac solution} has crossed terms involving right and left modes, it is much easier to calculate such averages using the $\chi$'s variables. Thus, in terms of $\chi_{1,k}$ and $\chi_{2,k}$ we have 
\begin{align}
    \braket{\mathcal{O}(t)}_{dBB}=\int D\chi_k\left|\Psi_{k}[\chi_{1,k},\chi_{1,k}^{*},\chi_{2,k},\chi_{2,k}^{*},\tau]\right|^2\mathcal{O}(\chi_{1,k},\chi_{1,k}^{*},\chi_{2,k},\chi_{2,k}^{*},\tau),
\end{align}
where we define the measure as $D\chi_k=d\chi_{1,k}d\chi_{1,k}^{*}d\chi_{2,k}d\chi_{2,k}^{*}$.

Taking into account that we can write $|\Psi_k|^2$ as
\begin{align}
    \left|\Psi_k\right|^2 =\frac{k^2}{\pi^2}\Re[H_{1,k}(\tau)]\Re[H_{2,k}(\tau)]e^{-2k\Re[H_{1,k}(\tau)]|\chi_{1,k}|^2}e^{-2k\Re[H_{2,k}(\tau)]|\chi_{2,k}|^2},
\end{align}
the effective mean values, for each wave number, of the expressions \eqref{kin ext}, \eqref{Vk ext}, and \eqref{Qk ext}  become, respectively,
\begin{align}
    \braket{K_k}_{dBB}&=\frac{k}{4}\left(\frac{\Im^2[H_{1,k}(\tau)]}{\Re{[H_{1,k}]}}+\frac{\Im^2[H_{2,k}(\tau)]}{\Re{[H_{2,k}]}}\right)=\frac{k \coth \left(\frac{\pi  k}{a}\right) \sin ^2(2 k t)}{\cosh \left(\frac{2 \pi  k}{a}\right)-\cos (4 k t)},\label{Kkmean ext}\\
    \braket{V_k}_{dBB}&=\frac{k}{4}\left(\frac{1}{\Re{[H_{1,k}(\tau)]}} +\frac{1}{\Re{[H_{2,k}(\tau)]}}\right)=\dfrac{k}{2} \coth \left(\frac{\pi  k}{a}\right),\label{Vkmean ext}\\
    \braket{Q_k}_{dBB}&=\frac{k}{4}\left(\Re{[H_{1,k}(\tau)]} +\Re{[H_{2,k}(\tau)]}\right)=\frac{k \sinh \left(\frac{2 \pi  k}{a}\right)}{2\left[\cosh \left(\frac{2 \pi  k}{a}\right)-\cos (4 k t)\right]}\label{Qkmean ext}.    
\end{align}

The average energy, which is the sum of the three terms above, reads,
\begin{align}\label{meanEnergy ext}
    \braket{E_k}_{dBB}=k\coth \left(\frac{\pi  k}{a}\right)=2k\left(\frac{1}{2}+\frac{1}{e^{\frac{2\pi}{a}k}-1}\right),
\end{align}
which is twice the energy when we consider just the right part, being consistent with the fact that each wedge should contribute with the same amount of energy. Interestingly, for high accelerations, which is equivalent to taking the limit of high temperatures, the effective mean energy is $2T$. Therefore, this result is in concordance with the equipartition theorem, which states that each quadratic term in Hamiltonian provides $T/2$ for the mean energy. Note that, in this case, the average value of the total classical potential is also time-independent, and half of the total average energy.

As in Section (\ref{KG}), we can obtain the mean number of Rindler particles in the Minkowski vacuum using the Hamiltonian operator ${\hat{H}}_k=\left(2{\hat{n}}_k+1\right)k$, as we have two massless non interacting scalar fields. Taking the average on both sides we have that $\braket{n_k}_{dBB}=(\frac{1}{k}\braket{E_k}_{dBB}-1)/2$, yielding, 
\begin{align}\label{nkext}
    \braket{n_k}_{dBB}=\frac{1}{e^{\frac{2\pi}{a}k}-1},
\end{align}
which is the Bose-Einstein distribution with Unruh temperature $T=a/2\pi$ for each one of the modes $\chi_{A,k}$. 

\begin{figure}[ht]
     \centering
     \begin{subfigure}[b]{0.4\textwidth}
         \centering
         \includegraphics[width=0.96\textwidth]{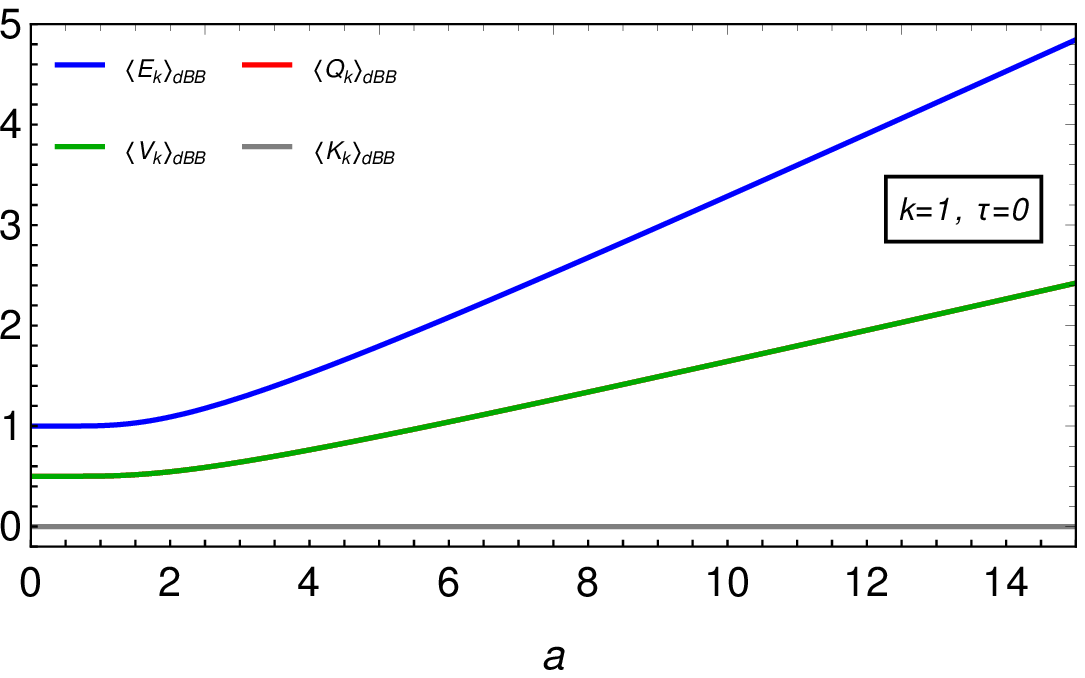}
         \caption{$k=1.0$ and $\tau=0$}
         \label{E_vs_a_t02}
     \end{subfigure}
     \centering
\begin{subfigure}[b]{0.4\textwidth}
         \centering
         \includegraphics[width=0.95\textwidth]{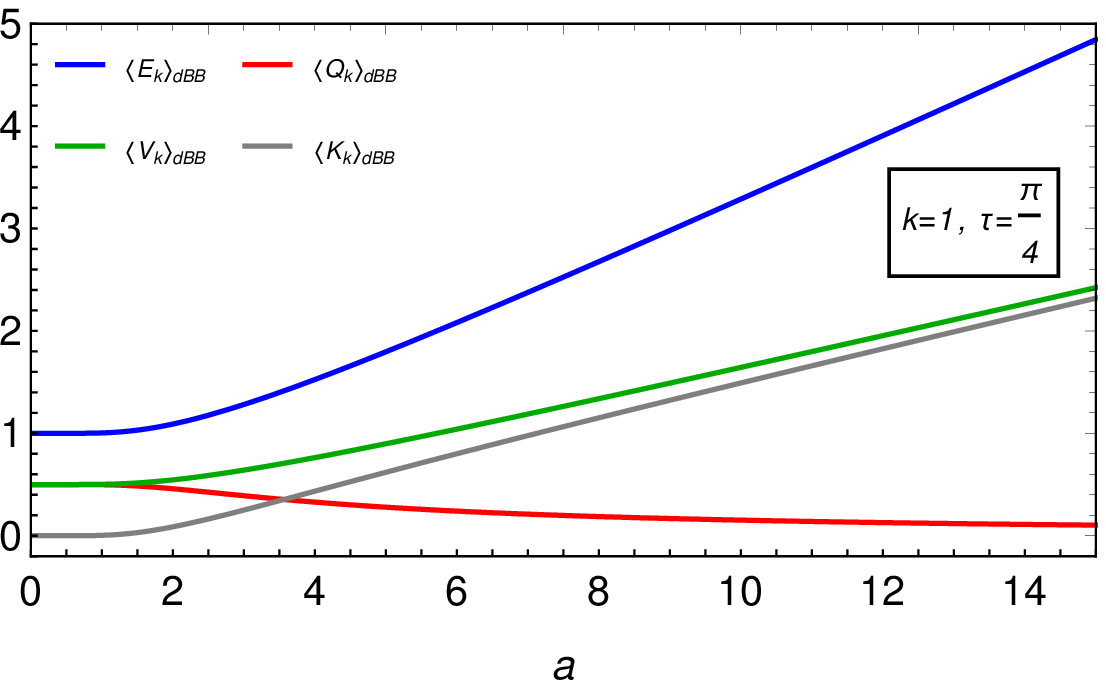}
         \caption{$k=1.0$ and $\tau=\pi/4$}
         \label{E_vs_a_t12}
\end{subfigure}
\caption{The mean values as functions of the parameter $a$ for (\ref{E_vs_a_t02}) $\tau=0$ and (\ref{E_vs_a_t12}) $\tau=\pi/4$. Note that for $\tau=0$, $\braket{Q_k}_{dBB}=\braket{V_k}_{dBB}$.}
\label{E_vs_a2}
\end{figure}

In Figure (\ref{E_vs_a2}) we analyze the behavior of the mean values as we increase the acceleration. For $\tau=0$ (Figure (\ref{E_vs_a_t02})) quantum and classical contributions are equivalents, since $\braket{Q_k}_{dBB}=\braket{V_k}_{dBB}$. For $\tau=\pi/4$ (Figure (\ref{E_vs_a_t12})), this equality is valid only for low values of $a$, with $\braket{Q_k}_{dBB}$ dropping to zero as the acceleration grows.

Let us now show the limits of the average parts of the total energy for low and high temperatures:

\subsection*{Low temperature (acceleration) regime: $T<<1$}

In this regime we have:

\begin{eqnarray}
\braket{K_k}_{dBB}&\approx& 2 k\,{\sin}^2(2k\tau)e^{-k/T}\approx 0,\\
\braket{V_k}_{dBB}&\approx& \frac{k}{2}+k e^{-k/T} \approx \frac{k}{2},\\
\braket{Q_k}_{dBB}&\approx& \frac{k}{2}+k\cos(4k\tau)e^{-k/T} \approx \frac{k}{2}.
\end{eqnarray}
Again, we recover the usual dBB picture of the Minkowski vacuum state: the energy of the field is equally shared between the classical and quantum potential, with negligible kinetic energy.

\subsection*{High temperature (acceleration) regime: $T>>1$}

There are two different situations:

\vspace{0.5cm}
\subsection*{i) $\boldsymbol{\tau \neq n\pi/(2k)}$, with $\boldsymbol{n}$ a an integer}
\vspace{0.5cm}

The results are:
\begin{eqnarray}
\braket{K_k}_{dBB}&\approx& T \\
\braket{V_k}_{dBB}&\approx& T \\
\braket{Q_k}_{dBB}&\approx& \frac{k^2}{4 T {\sin}^2(2 k\tau)}\approx 0
\end{eqnarray}
In this limit the classical kinetic and potential energies supply all the total energy $T$, with a negligible contribution of the quantum potential.

\vspace{0.5cm}
\subsection*{ii) $\boldsymbol{\tau = n\pi/(2k)}$, with $\boldsymbol{n}$ an integer.}
\vspace{0.5cm}

For these specific time values,
\begin{eqnarray}
\braket{K_k}_{dBB}&\approx& 0\\
\braket{V_k}_{dBB}&\approx& T\\
\braket{Q_k}_{dBB}&\approx& T 
\end{eqnarray}

As mentioned above, the total mean classical potential is time-independent and always contributes half of the total mean energy. The other half is now supplied by the mean kinetic energy, with abrupt shifts to the dominance of the mean quantum potential around $\tau = n\pi/(2k)$. Therefore, in the two-wedges case, the periodic spikes involve only the mean kinetic and quantum potential energies, interchanging half of the total energy.  

Let us now comment on the behavior of the mean values concerning only each of the fields $\chi_{1,k}$ and $\chi_{2,k}$ separately. The function appearing in the part of the mode wave function \eqref{vac solution xi} corresponding to the second field $\chi_{2,k}$, which is $H_{2,k}$, is the same as the one appearing in the mode wave function for $\phi_k^R$ (see Eqs.~(\ref{Psi radial},\ref{Re fk}) of the previous section), hence the $\chi_{2,k}$ contribution to the total energy and its parts have the same behavior as before. This means that this mode, as happens in the non-extended case, is responsible for the sudden spikes at $\tau=n\pi/k$ characteristic of the transition between classical and quantum dominance, explicit at high temperatures. For the $\chi_{1,k}$ field, however, its associated function $H_{1,k}$ can be obtained from $H_{2,k}$ by replacing $\cosh\left(\frac{\pi k}{a}\right)-\cos(2k\tau)$ with $\cosh\left(\frac{\pi k}{a}\right)+\cos(2k\tau)$, see Eqs.~(\ref{Re H1k},\ref{Re H2k}). Thus, the properties are similar but the jumps between classical and quantum dominance for large $T$ due to $\chi_{1,k}$ happen in the neighborhood of $\tau=\left(n+\frac{1}{2}\right)\pi/k$, with $n$ an integer.

As is the right-wedge case, this can also be seen from Eqs.~(\ref{KG bohmian ext R},\ref{KG bohmian ext L}). The field $\chi_{2,k}$ satisfies Eq.~\eqref{KG bohmian ext L}, which is identical to Eq.~\eqref{KG bohmian2}, hence yielding the same limits for (\ref{quantum force1},\ref{quantum force2}) of the previous section. In the case of the field $\chi_{1,k}$, however, satisfying Eq.~\eqref{KG bohmian ext R}, the limits in the high temperature regime, $T>>1$, are:

\begin{align}\label{quantum force3}
    \frac{\partial^2 \chi_{1,k}}{\partial \tau^2} + k^2\chi_{1,k}
    \approx \frac{k^4}{16 T^2 {\cos}^2(k\tau)}\chi_{1,k}\approx 0\; ;\;\;\;\;\tau\neq \left(n+\frac{1}{2}\right)\frac{\pi}{k},
\end{align}

\begin{align}
    \frac{\partial^2 \chi_{1,k}}{\partial \tau^2} + k^2\chi_{1,k} \approx 16T^{2}\chi_{1,k}\; ;\;\;\;\;\tau =\left(n+\frac{1}{2}\right)\frac{\pi}{k}\label{quantum force4},
\end{align}
The Bohmian field $\chi_{1,k}$ obeys a classical Klein-Gordon equation when $\tau\neq \left(n+\frac{1}{2}\right)\pi/k$, as the quantum force is negligible. However, in the vicinity of $\tau=\left(n+\frac{1}{2}\right)\pi/k$, it becomes the dominant force, resulting in a sudden transition from classical to quantum dominance.

\begin{figure}[ht]
     \centering
     \includegraphics[width=0.55\textwidth]{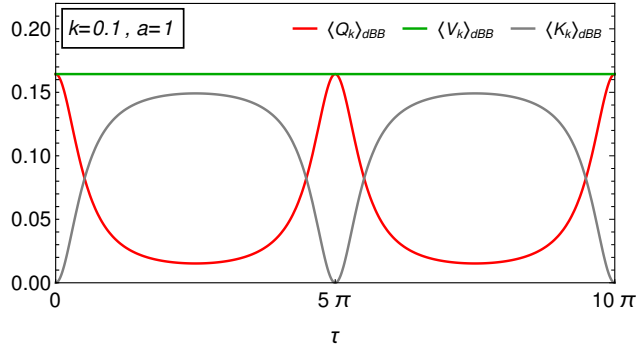}
     \caption{Mean values of the kinetic term and the quantum potential for the extended case, with $k=0.1$ and $a=1$. The sum of $\braket{Q_k}_{dBB}$ and $\braket{K_k}_{dBB}$ is exactly the average value of the classical potential $\braket{V_k}_{dBB}$, represented as a straight line.}
     \label{a1ext}
\end{figure}

\begin{figure}[ht]
     \centering
     \includegraphics[width=0.55\textwidth]{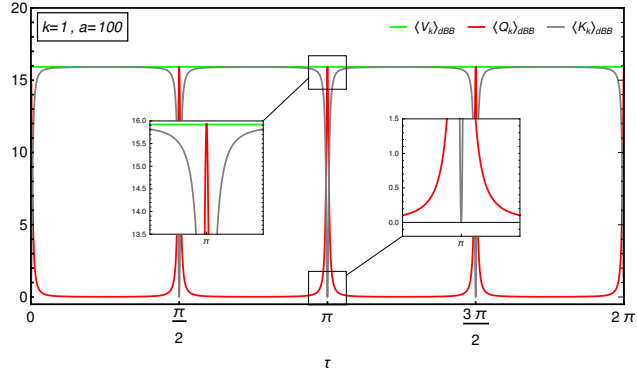}
     \caption{Representative plot of $\braket{Q_k}_{dBB}$ and $\braket{K_k}_{dBB}$ with their sum, $\braket{V_k}_{dBB}$, for $k=1$ and $a=10^2$. Near $\tau=n\pi/(2k)$, the kinetic contribution sharply shifts to the quantum potential, which quickly dominates as the kinetic term drops to zero.}
     \label{a100ext}
\end{figure}

In Figure (\ref{a1ext}) we plot all mean energies for accelerations of order $1$. In Figure (\ref{a100ext}) we plot the mean kinetic energy together with the mean quantum potential energy and their sum, for the case of high accelerations (temperatures): $a=10^2$. It is evident that near $\tau = n\pi/(2k)$ the quantum potential and the kinetic term switch their roles, with $\braket{K_k}_{dBB}$ dropping to zero as $\braket{Q_k}_{dBB}$ grows. For large accelerations, this behavior is characterized by sharp peaks centered at $\tau = n\pi/(2k)$.

\subsection{Extended field trajectories}

Similarly to the non-extended case we have a special field configuration, solution to dBB guidance equations, with analogous properties of the average values computed in the last subsection. From Eqs. \eqref{guidance eq 1 extended xi} and \eqref{guidance eq 2 extended xi} we obtain
\begin{align}
    \chi_{1,k}(\tau)&=\frac{D_{1,k}}{\sqrt{2k\Re[H_{1,k}(\tau)]}},\hspace{6.0mm}
    \chi_{2,k}(\tau)=\frac{D_{2,k}}{\sqrt{2k\Re[H_{2,k}(\tau)]}}.
    \label{guidanc sol chi1} 
\end{align}
The probability density distribution takes a very simple form, the Gaussian $|\Psi_k|^2\propto e^{-|D_{1,k}(a)|^2-|D_{2,k}(a)|^2}$. As in 
Sec.~(\ref{KG}), fields with $|D_{1,k}|=|D_{2,k}|=1$ are the unique possibility of Bohmian fields with time-independent energy 
\eqref{Ek ext}, which is equal to the average energy given in equation \eqref{meanEnergy ext}. Moreover, each individual part of the total energy of such Bohmian fields is equal to its own average, namely, $Q_k=\braket{Q_k}_{dBB}$, $V_k=\braket{V_k}_{dBB}$, and $K_k=\braket{K_k}_{dBB}$. Hence, the asymptotic limits of the average quantities calculated in the previous subsection are also valid for every single Bohmian field with 
$D_{1,k}(a)=\exp(i \theta_{1,k}(a)), D_{2,k}(a)=\exp(i \theta_{2,k}(a))$, including the periodic abrupt shifts from classical kinetic to quantum potential dominance discussed above.

The asymptotic behaviors of these particular Bohmian fields, disregarding their phase, read

\begin{align}\label{Bohmian field1} 
\chi_{1,k}&=\frac{1+\cos(2 k \tau) e^{-k/(2T)}}{\sqrt{2k}}\; , \;\;\;\;\; T<<1\\
\chi_{1,k}&=\frac{\sqrt{2T}|\cos(k \tau)|}{k}\; , \;\;\;\;\; T>>1
\end{align}

\begin{align}\label{Bohmian field2} 
\chi_{2,k}&=\frac{1-\cos(2 k \tau) e^{-k/(2T)}}{\sqrt{2k}}\; , \;\;\;\;\; T<<1\\
\chi_{2,k}&=\frac{\sqrt{2T}|\sin(k \tau)|}{k}\; . \;\;\;\;\; T>>1
\end{align}

\subsection{Power Spectrum for the complete manifold}

As in the previous section, we would like to obtain the power spectrum for the associated right and left modes. In the case of the two-wedge problem, it is defined as
\begin{align}
    (P^{ab})_k(\tau)=\int_{-\infty}^{\infty}d\xi e^{-ik\xi}\left <\phi^a(\xi)\phi^b(0)\right>_{dBB}, \label{power spectrum extended}
\end{align}
where $\phi^a$ is the inverse Fourier transform of $\phi^a_k$, with $a,b=R,L$, while the power spectrum for variables $\chi_{A,k}$ is
\begin{align}
    (P_{AB})_k(\tau)=\int_{-\infty}^{\infty}d\xi e^{-ik\xi}\left <\chi_A(\xi)\chi_B(0)\right>_{dBB}, \label{power spectrum extended xi}
\end{align}
with $\chi_A$ being the inverse Fourier transform of $\chi_{A,k}$, $A,B=1,2$.
The calculation of the correlations among the $\chi_A$ modes reveals that
\begin{align}\label{two-p f xi}
    \left<\chi_{A}(\xi)\chi_{B}(0)\right>_{dBB}=\frac{1}{2\pi}\int_{-\infty}^{\infty} dke^{i k \xi}\frac{\delta_{AB}}{2|k| \Re[H_{A,k}(\tau)]},
\end{align}
with a null crossed correlation. The nonzero components of the associated power spectrum are
\begin{align}
    (P_{11})_k(\tau)=&\frac{1}{2k\Re[H_{1,k(\tau)}]}=\frac{ \cosh \left(\frac{\pi  k}{a}\right)+\cos (2 k t)}{2 k \sinh\left(\frac{\pi k}{a}\right)},\label{P11}\\
    (P_{22})_k(\tau)=&\frac{1}{2k\Re[H_{2,k(\tau)}]}=\frac{ \cosh \left(\frac{\pi  k}{a}\right)-\cos (2 k t)}{2 k \sinh\left(\frac{\pi k}{a}\right)}. \label{P22}
\end{align}
So, for high temperatures, we have that
\begin{align}
    (P_{11})_k(\tau)\simeq\frac{2T}{k^2}\cos^2(k\tau), \hspace{6.0mm} (P_{22})_k(\tau)\simeq\frac{2T}{k^2}\sin^2(k\tau),
\end{align}
while for low temperatures 
\begin{align}
    (P_{11})_k(\tau)\simeq (P_{22})_k(\tau)\simeq\frac{1}{2k}.
\end{align}
Such results are very similar to those found in the non-extended case, being a consequence of the fact that the wave functional \eqref{vac solution xi} behaves like two independent Minkowski ground states. As a matter of fact, $(P^{11})_k$ and $(P^{22})_k$ are related with the respective contribution to the classical potential due to $\chi_{1,k}$ and $\chi_{2,k}$, that is to say
\begin{align}
    (P_{11})_k(\tau)&=\frac{2}{k^2}\braket{V_{1,k}}_{dBB},\\
    (P_{22})_k(\tau)&=\frac{2}{k^2}\braket{V_{2,k}}_{dBB},
\end{align}
where $V_{1,k}=\frac{1}{2}k^2|\chi_{1,k}|^2$ and $V_{2,k}=\frac{1}{2}k^2|\chi_{2,k}|^2$.

It is possible to express the original correlations $\left <\phi^a(\xi)\phi^b(0)\right>_{dBB}$ in terms of Eq. \eqref{two-p f xi} such that the power spectrum \eqref{power spectrum extended} becomes
\begin{align}
    (P^{RR})_k(\tau)=(P^{LL})_k(\tau)&=\frac{1}{4k}\left(\frac{1}{\Re[H_{1,k}(\tau)]}+\frac{1}{\Re[H_{2,k}(\tau)]}\right)=\frac{\coth\left(\frac{\pi k}{a}\right)}{2k},
\end{align}
which can be related to the classical potential as follows  
\begin{align}
    (P^{RR})_k(\tau)=(P^{LL})_k(\tau)=\frac{1}{k^2}\braket{V_{k}}_{dBB}.
\end{align}
Conversely, 
\begin{align}
    (P^{RL})_k(\tau)=(P^{LR})_k(\tau)&=\frac{1}{4k}\left(\frac{1}{\Re[H_{1,k}(\tau)]}-\frac{1}{\Re[H_{2,k}(\tau)]}\right)=\frac{\cos(2k\tau)}{2k\sinh\left(\frac{\pi k}{a}\right)},
\end{align}
indicating a non-null correlation between the right and left modes. In the high-temperature limit
\begin{align}\label{PWrl}
 (P^{RR})_k(\tau)=(P^{LL})_k(\tau)\simeq\frac{T}{k^2}, \hspace{6.0mm} (P^{RL})_k(\tau)=(P^{LR})_k(\tau)\simeq\frac{T}{k^2}\cos(2k\tau).
\end{align}
Note that in the common spacelike hypersurfaces $\tau=t=0$, the results obtained above are identical to the power spectrum of a classical field at finite temperature in Minkowski space, see Ref.~\cite{Aarts:1997kp}. 

For low accelerations, we have that 
\begin{align}
    (P^{RR})_k(\tau)=(P^{LL})_k(\tau)\simeq\frac{1}{2k},\hspace{6.0mm} (P^{RL})_k(\tau)=(P^{LR})_k(\tau)\simeq0.
\end{align} 
indicating that, in this case, the crossed correlations are negligible. Thus, if the effect of the horizon is not evident, the correspondent non-local connection between the left and right wedges can be neglected.

\section{Conclusions} 

In this paper, we analyzed the behavior of a massless scalar field in the Rindler spacetime from the de Broglie-Bohm (dBB) perspective. Our study aimed to understand Bohmian aspects of the Unruh effect by considering first the right Rindler wedge, and then extending our analysis to include the left side as well. In both cases, we obtained a Hamilton-Jacobi-like equation for the Bohmian fields, together with their guidance equations, recovering the known results of a Bohmian field in the Minkowski vacuum for low accelerations. 

Using the dBB techniques for arbitrary accelerations, we calculated the average energy, obtaining the Bose-Einstein distribution with Unruh temperature for the mean value of the total energy. As the distribution of initial field configurations satisfies the Born rule, the final result obtained using the dBB approach must be exactly the same as the one using standard techniques. Therefore, at first glance, there is nothing new. However, by using the Hamilton-Jacobi-like equation for the Bohmian fields, the dBB approach offers a different perspective on the phenomenon, as it allows the separation of the total mean energy into classical and quantum parts, which is not possible with the standard approach. Inspecting these terms, we observed a periodic interchange between quantum and classical contributions as the leading cause of temperature-associated effects, more prominent for large accelerations. More precisely, for $a/k >> 1$, which can also be viewed as an infrared limit, this quantum-classical alternation presents highly abrupt jumps around $\tau = n\pi/(2k)$, where $n$ an integer. We do not know if these effects can be observed. Note that, assuming the Born rule, the statistical predictions of the dBB quantum theory are the same as in the usual approach. However, regarding a quantum phenomenon from a different perspective can help in the search for new experimental consequences, which would be very hard to be seen using the standard point of view. In the case of the dBB quantum theory, there are many concrete examples of this assertion, see Ref.~\cite{StruyvePinto} for details. In the present case, the sudden transitions between classical and quantum dominance mentioned above do not appear to be simple artefacts of the dBB approach, as long as such abrupt jumps also appear in the mode wave function solutions themselves, see Eqs.~(\ref{Psi radial},\ref{vac solution}) for $a/k >> 1$ around $\tau = n\pi/k$ for the RH-wedge and $\tau = n\pi/(2k)$ for the extended case. In fact, they seem to be manifestations of the jumps at the wave functional level, which may indeed lead to experimental consequences.

We solved the guidance equations and found a very peculiar Bohmian field configuration in which its individual total energy, classical potential, classical kinetic energy, and quantum potential were all exactly equal to their corresponding mean values, with the emergence of an effective Unruh temperature. We would like to emphasize that this is the unique Bohmian field solution with time-independent energy. Note that the Unruh temperature appears even within an individual field configuration, making it not only an averaged property of the quantum state.

We have seen that the Bohmian field in the Rindler frame obeys an effective Klein-Gordon equation with an effective mass that depends on the temperature, see Eqs~(\ref{quantum force1},\ref{quantum force2}) and (\ref{quantum force3},\ref{quantum force4}). As they mimic a quantum field, they can perhaps be explored to construct analog models of the Unruh effect. 

In the case of the complete manifold analysis, we have seen the non-local nature of the guidance equations (\ref{guidance eq 1 extended},\ref{guidance eq 2 extended}) for the Bohmian field modes defined in the right and left wedges: the dynamics of the right (left) mode is affected by the left (right) mode, even though they are separated by a horizon. This can be useful to understand better the entanglement between these two field modes, and to possibly extract some physical consequences from it, opening the way for a black hole analysis.

Finally, as a last speculation, we have commented that the dBB approach can lead to different results from the standard quantum theory for some period of time, before reaching quantum equilibrium, when the distribution of initial field configurations is not given by the Born rule. Hence, taking the ensemble of field configurations given in Eqs.~(\ref{bohmian field},\ref{guidanc sol chi1}), and taking the distributions of the integration constants $D_{A,k}(a)$ different from $|\Psi_k|^2$ at some initial time, it would be interesting to investigate what kind of particle distribution would emerge, its associated temperature, if it exists, and how long it would take to reach the quantum equilibrium. In Ref.~\cite{Valentini1} it is argued that quantum black holes can violate the Born rule, with consequences to the Hawking radiation. The simple model studied here can be a point of departure to investigate this possibility more precisely.

\section*{Acknowledgments}
The authors would like to thank and acknowledge financial support
from the National Scientific and Technological Research Council (CNPq,
Brazil). NPN acknowledges
the support of CNPq of Brazil under grant PQ-IB
310121/2021-3. The authors are also grateful to Prof. Sebastião Alves Dias for some illuminating discussions on this problem.

\appendix
\section*{Appendix A}

In this appendix, we write the explicit expressions and limits used throughout the text. 

\subsection*{Right-Rindler wedge}\label{app2}

The energy obtained via the Hamilton-Jacobi equation \eqref{real schrod} through the equation \eqref{Ek} is, in terms of $\phi_k^R$, such that
\begin{align}
    E_k(\tau)=\frac{ 1-\cosh \left(\frac{\pi  k}{a}\right) \cos (2 k t)}{\left[\cosh \left(\frac{\pi  k}{a}\right)-\cos (2 k t)\right]^2}k^2|\phi_k^R|^2+\frac{k \sinh \left(\frac{\pi  k}{a}\right)}{2\left[\cosh \left(\frac{\pi  k}{a}\right)-\cos (2 k t)\right]},\nonumber
\end{align}
while the other contributions are
\begin{align*}
    Q_k(\tau)&=\frac{\sinh \left(\frac{\pi  k}{a}\right)}{2\left[\cosh \left(\frac{\pi  k}{a}\right)-\cos (2 k t)\right]}k-\frac{\sinh ^2\left(\frac{\pi  k}{a}\right)}{2\left[\cosh \left(\frac{\pi  k}{a}\right)-\cos (2 k t)\right]^2}k^2|\phi_k^R|^2,\\
    V_k(\tau)&=\frac{1}{2}k^2|\phi_k^R|^2,\\
    K_k(\tau)&=\frac{\sin ^2(2 k t)}{2\left[\cosh \left(\frac{\pi  k}{a}\right)-\cos (2 k t)\right]^2}k^2|\phi_k^R|^2.
\end{align*}

\subsubsection*{Low temperatures}
For $\frac{\pi k}{a}\gg1$, we use the fact that
\begingroup
\setlength{\tabcolsep}{10pt}
\renewcommand{\arraystretch}{2.2} 
\begin{align*}
\begin{tabular}{ l l }
$\Re[f_k(\tau)]\approx1,$ & $\Im[f_k(\tau)]\approx0,$\\
$\Re[\Omega_k(\tau)]\approx-\dfrac{k}{4T},$ & $\Im[\Omega_k(\tau)]\approx-k\tau.$\\
\end{tabular}
\end{align*}
\endgroup
Therefore, the wave functional \eqref{Psi radial} can be expressed as
\begin{align}
    \Psi_k[\phi_k^R,\phi_k^{R*},\tau]\approx\sqrt{\frac{k}{\pi}}e^{-k|\phi_k^R|^2-ik\tau}.\nonumber
\end{align}

\subsubsection*{High temperatures}
1) $\tau\ne\dfrac{n \pi}{k}$, with $n$ an integer\vspace{0.5cm}

When $\dfrac{\pi k}{a}\ll1$, we have the following expansions for the coefficients 
\begingroup
\setlength{\tabcolsep}{10pt}
\renewcommand{\arraystretch}{2.2} 
\begin{align*}
\begin{tabular}{ l l }
$\Re[f_k(\tau)]\approx\dfrac{k}{4T\sin^2(k\tau)},$ & $\Im[f_k(\tau)]\approx-\cot(k\tau),$\\
$\Re[\Omega_k(\tau)]\approx\ln\sqrt{2}-\dfrac{1}{2}\ln\left(1-\cos(2k\tau)\right),$ & $\Im[\Omega_k(\tau)]\approx-\dfrac{\pi}{2}\sign\left(\tan(k\tau)\right).$\\
\end{tabular}
\end{align*}
\endgroup
Therefore, the wave functional is
\begin{align}
    \Psi_k[\phi_k^R,\phi_k^{R*},\tau]\approx&\frac{k}{\sqrt{4\pi T}}\frac{1}{|\sin(k\tau)|}\exp{\left\{-\frac{k^2}{4T\sin^2(2k\tau)}|\phi_k^R|^2\right\}}\times\nonumber\\
    &\exp{\left\{ik\cot(k\tau)|\phi_k^R|^2-i\frac{\pi}{2}\sign(\tan(k\tau))\right\}},\nonumber
\end{align}

2) $\tau=\dfrac{n \pi}{k}$, with $n$ an integer\vspace{0.5cm}

For these specific times, the quantum potential is the dominant contribution in the Hamilton-Jacobi equation \eqref{real schrod}. In this case
\begingroup
\setlength{\tabcolsep}{10pt}
\renewcommand{\arraystretch}{2.2} 
\begin{align*}
\begin{tabular}{ l l }
$\Re[f_k(\tau)]\approx\dfrac{4T}{k},$ & $\Im[f_k(\tau)]=0,$\\
$\Re[\Omega_k(\tau)]\approx\ln\left(\dfrac{4T}{k}\right),$ & $\Im[\Omega_k(\tau)]=0.$\\
\end{tabular}
\end{align*}
\endgroup

with the wave functional
\begin{align}
    \Psi_k[\phi_k^R,\phi_k^{R*},\tau]\approx&\sqrt{\frac{4T}{\pi}}\exp{\left\{-4T|\phi_k^R|^2\right\}}.\nonumber
\end{align}

\subsection*{Extended case}\label{app3}

The explicit expression for the energy in terms of the field modes $\chi_{1,k}$ and $\chi_{2,k}$ is
\begin{align*}
    E_k(\tau) = & k^2\frac{1+\cosh \left(\frac{\pi  k}{a}\right) \cos (2 k t)}{\left[\cosh \left(\frac{\pi  k}{a}\right)+\cos (2 k t)\right]^2}|\chi_{1,k}|^2+ k^2\frac{1-\cosh \left(\frac{\pi  k}{a}\right) \cos (2 k t)}{\left[\cosh \left(\frac{\pi  k}{a}\right)-\cos (2 k t)\right]^2}|\chi_{2,k}|^2+\\
    & k\frac{ \sinh \left(\frac{2 \pi  k}{a}\right)}{\cosh \left(\frac{2 \pi  k}{a}\right)-\cos (4 k t)}.
\end{align*}

Conversely, the quantum and classical potentials, together with the kinetic contribution in the Hamilton-Jacobi equation \eqref{real schrod ext}, are, respectively
\begin{align*}
    Q_k(\tau)&=-\frac{\sinh ^2\left(\frac{\pi  k}{a}\right)}{2\left[\cosh \left(\frac{\pi  k}{a}\right)+\cos (2 k t)\right]^2}k^2|\chi_{1,k}|^2-\frac{\sinh ^2\left(\frac{\pi  k}{a}\right)}{2\left[\cosh \left(\frac{\pi  k}{a}\right)-\cos (2 k t)\right]^2}k^2|\chi_{2,k}|^2+\\
    &\hspace{4.0mm}k\frac{\sinh \left(\frac{2\pi  k}{a}\right)}{\cosh \left(\frac{2\pi  k}{a}\right)-\cos (4 k t)}\\
    V_k(\tau)&=\frac{1}{2}k^2|\chi_{1,k}|^2+\frac{1}{2}k^2|\chi_{2,k}|^2,\\
    K_k(\tau)&=\frac{\sin^2\left(2k\tau\right)}{2\left[\cosh \left(\frac{\pi  k}{a}\right)+\cos (2 k t)\right]^2}k^2|\chi_{1,k}|^2+\frac{\sin ^2\left(2k\tau\right)}{2\left[\cosh \left(\frac{\pi  k}{a}\right)-\cos (2 k t)\right]^2}k^2|\chi_{2,k}|^2.\nonumber
\end{align*}

\subsubsection*{Low temperatures}
In the low-temperature regime we have that 
\begingroup
\setlength{\tabcolsep}{10pt}
\renewcommand{\arraystretch}{2.2} 
\begin{align*}
\begin{tabular}{ l l }
$\Re[H_{1,k}(\tau)]\approx\Re[H_{2,k}(\tau)]\approx1,$ & $\Im[H_{1,k}(\tau)]\approx\Im[H_{2,k}(\tau)]\approx0,$\\
$\Re[\Theta_k(\tau)]\approx-\dfrac{k}{2T},$ & $\Im[\Theta_k(\tau)]\approx-2k\tau.$\\
\end{tabular}
\end{align*}
\endgroup
So, the wave functional can be written as
\begin{align}
    \Psi_k[\chi,\tau]\approx\frac{k}{\pi}e^{-k|\chi_{1,k}|^2-k|\chi_{2,k}|^2-2ik\tau}\nonumber.
\end{align}

\subsubsection*{High temperatures}
1) $\tau\ne\dfrac{n \pi}{2k}$, with $n$ an integer\vspace{0.5cm}

The expansion of the coefficients for high temperatures is given by
\begingroup
\setlength{\tabcolsep}{10pt}
\renewcommand{\arraystretch}{2.2} 
\begin{align*}
\begin{tabular}{ l l }
$\Re[H_{1,k}(\tau)]\approx\dfrac{k}{4T\cos^2(k\tau)},$ & $\Im[H_{1,k}(\tau)]\approx\tan(k\tau),$\\
$\Re[H_{2,k}(\tau)]\approx\dfrac{k}{4T\sin^2(k\tau)},$ & $\Im[H_{2,k}(\tau)]\approx-\cot(k\tau),$\\
$\Re[\Omega_k(\tau)]\approx\ln\sqrt{2}-\dfrac{1}{2}\ln\left(1-\cos(4k\tau)\right),$ & $\Im[\Omega_k(\tau)]\approx-\dfrac{\pi}{2}\sign\left(\tan(2k\tau)\right).$\\
\end{tabular}
\end{align*}
\endgroup
Therefore the wave functional is
\begin{align}
    \Psi_k[\chi,\tau]\approx
    &\frac{k^2}{2\pi T}\frac{1}{|\sin(2k\tau)|}\exp{\left\{-\frac{k^2}{4T\cos^2(k\tau)}|\chi_{1,k}|^2-\frac{k^2}{4T\sin^2(k\tau)}|\chi_{2,k}|^2\right\}}\times\nonumber\\
    &\exp{\left\{i\left[-k\tan(k\tau)|\chi_{1,k}|^2+k\cot(k\tau)|\chi_{2,k}|^2-\dfrac{\pi}{2}\sign\left(\tan(k\tau)\right)\right]\right\}}.\nonumber
\end{align}

2) $\tau=\dfrac{n \pi}{2k}$, with $n$ an integer\vspace{0.5cm}

For these specific times, we have that
\begingroup
\setlength{\tabcolsep}{10pt}
\renewcommand{\arraystretch}{2.2} 
\begin{align*}
\begin{tabular}{ l l }
$\Re[H_{1,k}(\tau)]\approx\dfrac{k}{4T},$ & $\Im[H_{1,k}(\tau)]\approx 0,$\\
$\Re[H_{2,k}(\tau)]\approx\dfrac{4T}{k},$ & $\Im[H_{2,k}(\tau)]\approx 0,$\\
$\Re[\Omega_k(\tau)]\approx\ln\left(\dfrac{4T}{k}\right),$ & $\Im[\Omega_k(\tau)]\approx0.$\\
\end{tabular}
\end{align*}
\endgroup
So, the wave functional can be approximated by
\begin{align}
    \Psi_k[\chi,\tau]\approx&\frac{k}{\pi}\exp{\left\{-\frac{k^2}{4T}|\chi_{1,k}|^2-4T|\chi_{2,k}|^2\right\}}.\nonumber
\end{align}

\end{document}